\documentclass[pre,twocolumn,showpacs,floatfix]{revtex4-1}%

\usepackage[english]{babel}
\usepackage[utf8]{inputenc}
\usepackage{amssymb}
\usepackage{amsmath}
\usepackage{graphicx}
\usepackage[colorinlistoftodos]{todonotes}
\usepackage{xcolor}
\usepackage{float}

\UseRawInputEncoding

\usepackage{blindtext}

\usepackage{color}
\usepackage[normalem]{ulem}
\usepackage{mathtools}
\usepackage{color}
\usepackage{lineno} 
\usepackage{amsfonts}
\usepackage{soul}
\usepackage{bm}
\usepackage{float}

\begin{document}
\title{Emergence of Directed Motion in a Crowded Suspension of Overdamped Particles}
\author{Deborah Schwarcz$^1$}
\email{deborah.schwarcz@gmail.com}
\author{Stanislav Burov$^2$}
\email{stasbur@gmail.com}
\affiliation{$^1$Department of Mathematics and $^2$Physics Department, Bar-Ilan 
University, Ramat Gan 5290002,
Israel}


\begin{abstract}
 In this work, we focus on the behavior of a single passive Brownian particle in a suspension of passive particles with short-range repulsive interactions and a larger self-diffusion coefficient. While the forces affecting the single-particle are thermal-like fluctuations and repulsion, due to other particles in the suspension,  our numerical simulations show that on intermediate time scales directed motion on a single-particle level emerges. 
 This emergent directional motion leads to a breakdown of the Einstein relation and non-monotonic augmentation of the measured diffusion coefficient. 
 Directional tendency increases with the density of the suspension and leads to growth of the diffusivity with the density of the suspension, a phenomenon recently observed for a system of hard spheres by Ilker, Castellana, and Joanny~\cite{Joanny}. 
 Counter-intuitively, the directional flow originates from the tendency of different particles  to push each other out of their way. 
 Due to such strictly repulsive interactions, nearby particles form into temporally correlated pairs and move cooperatively, thus creating a preferred direction of motion on intermediate time scales.   
 We show that directional motion emerges when the ratio of the self-diffusion coefficients of the tracked particle and suspension constituents is below a critical value.
 \end{abstract}

\maketitle

\section{Introduction}
Active particles are defined by the ability to perform  directed motion while consuming energy from the environment and dissipating it back~\cite{RevModPhys2016,Ramaswamy2010,Cates2012,Self_propelled_particle}. 
This signature of directional transport of active particles contrasts with the passive motion of Brownian particles, which are driven by thermal fluctuations. 
Active matter, like the cytoskeleton, relies on the presence of systematic motion of the basic ingredients, e.g., molecular motors ~\cite{Marchetti2013}. 
The presence of directionality in the movement can be exploited; for example, bacteria's run and tumble property was utilized to operate micro machines~\cite{Aranson2010,Fabrizio2010}. Heat engines that are designed to use fluctuations on a single  particle level~\cite{Bechinger2012,Dechant2015,Rica2016} experience an increase in efficiency when the surrounding reservoir includes active ingredients~\cite{Sood2016}. 
While active particles are naturally present in the biological world, e.g., bacteria and molecular motors, numerous artificial examples  are based on chemical reactions~\cite{RevModPhys2016} or manipulation by external fields~\cite{Norbert2014}.
For example, directed propulsion can be achieved for Janus spherical particles by unequally coating the surface with platinum~\cite{Golestanian2007}. 
In this work we focus on appearance of directional motion for overdamped passive Brownian particles.

The appearance of macroscopic and asymmetric space-time states (namely directional flows) 
was the focus of the dissipative structures theory~\cite{Prigogine1978}. 
The most frequently used example of a dissipative structure is the Rayleigh-B\'{e}nard instability~\cite{Kadanoff2001} where macroscopic convection starts when a critical temperature difference of the confining plates is reached. 
The existence of a critical point, beyond which a novel  non-equilibrium dynamic state is formed, is the sole idea behind dissipative structures on the macro-scale. 
Is such a critical point exists 
when a single-particle (SP) is considered? 
Can directed motion of a SP, be an emergent phenomenon when a passive Brownian particle (PBP) is pushed far enough into the non-equilibrium realm of a thermal system? 
In  theoretical studies of binary mixtures of PBPs with different diffusivites~\cite{Grosberg2015,Weber2016}, i.e., temperatures, a demixing transition (for sufficiently high concentration) was observed. 
Such demixing into phases with low and high diffusivities resembles the motility-induced phase separation  to slow and fast active particles present in active systems~\cite{Cates2016}. 
Moreover, a system consisting of two particles in contact with different thermostats and a binding potential (a model that is easily mapped to a problem of two different diffusivities ) reaches a steady-state that doesn't satisfy Boltzmann statistics~\cite{Grosberg2015,Oshanin2013,Oshanin2019,Grosberg2018}. 
The non-equilibrium properties of two PBPs in a contact with different thermostats, and pair-wise quadratic interactions, were utilized  to identify broken detailed balance on a mesoscopic scale~\cite{Battle2016} and to quantify dissipation~\cite{ref_Nat_Comm}.
Non-equilibrium steady states were also studied when three PBPs with different temperatures are considered~\cite{different_temp_Grosberg}. It was shown that PBPs with higher diffusivity (``hot" particles) effectively attract each other when immersed in a suspension of PBPs with lower diffusivity (``cold" particles)~\cite{PhysRevFluids_harvard}. 
Moreover, the effective temperature of a ``hot" tracer decreases due to interaction with a bath of ``cold" particles~\cite{Grosberg_2021}.
Recently, it was found  that the long-time diffusivity of a single ``cold" PBP in a solution of ``hot" PBPs, increases as a function of ``hot" PBPs self-diffusivity and density \cite{Joanny,soft_particles_different_thermostats}.  

In this article, we explore the microscopic mechanism that leads to the enhancement of the 
long-time diffusion coefficient of a ``cold" PBP in a bath of ``hot" PBPs. 
We observe that 
once the diffusivities/temperature ratio of the "cold" and ``hot" PBPs crosses a specific value,
a presence of directional motion appears in the transport of the ``cold" PBP. 
The time scales when the "cold" PBP moves in a directed fashion depend on the density of the surrounding particles. 
Our numerical simulations show a rich behavior of the tracked SP. 
We reveal critical transition to directed motion, breakdown of the Einstein relation, and non-monotonic behavior of the long-time diffusion coefficient as a function of the  density of the solution. 

The article is organized as follows: In Sec.~\ref{secmodel} we present our model: a two-dimensional suspension of ``hot" PBPs and a single ``cold" PBP that is immersed into this suspension. In Section \ref{secmsd}, we analyze the mean square displacement of the ``cold" PBP and demonstrate the enhancement of the long-time diffusion coefficient. 
In Section \ref{secrelativeangle}, we utilize the behavior of the distribution of the relative angle of the ``cold" PBP to observe appearance of directed motion. 
In section \ref{sectiontcp} we show that the ``cold" PBP tends to interact with a single ``hot" PBP, i.e., nearest neighbor, forming a temporarily correlated pair.
Our findings strongly suggest that the spontaneous directional symmetry breaking of ``cold" PBPs is associated with the formation of such temporarily correlated pairs. 
The discussion and summary of our findings are provided in Sec.~\ref{secdiscussion}.





\section{The Model}
\label{secmodel}
We perform simulations of a  $2$-dimensional solution of PBPs. The position of every particle $i$, $\vec{r}_i$, is governed by the over-damped Langevin equation~\cite{Gardiner}
\begin{equation}
\frac{d \vec{r}_i}{dt} = \frac{1}{\gamma}\sum_j\vec{F}(r_{i,j}) + \sqrt{2 D_i}\vec{\eta}_i(t)
   \label{eq:langevinDef}
\end{equation}
where $\vec{F}(r_{i,j})$ is the force due to interaction with particle $j$ and $r_{i,j}=|\vec{r}_i-\vec{r}_j|$ is the distance between particle $i$ and $j$. $D_i$ is the diffusion coefficient of particle $i$ and $\vec{\eta}_i(t)=\eta_i^x\hat{x}+\eta_i^y\hat{y}$, while $\eta_i^x$ (and  $\eta_i^y$) is the random $\delta$-correlated Gaussian noise: i.e. $P(\eta_i^x)=\frac{1}{\sqrt{2\pi}}\exp\left(-(\eta_i^x)^2/2\right)$ and  $\langle \eta_i^x(t)\eta_j^x(t')=\delta_{i,j}\delta(t-t')$. Coefficient $1/\gamma$ plays the role of inverse friction coefficient, and in this work, we use $\gamma=1$.
The force $\vec{F}(r_{i,j})$ is determined by the Weeks-Chandler-Andersen (WCA potential)~\cite{Chandler_WCA} (the purely repulsive part of Lennard-Jones potential)
  \begin{equation}
  \label{eq_wcapotential}
  V_{i,j}=
    \begin{cases}
      4\epsilon \left[ \left(\sigma /r_{i,j} \right)^{12}-\left(\sigma /r_{i,j} \right)^{6}\right] &   r_{i,j}\leq2^{\frac{1}{6}}\sigma  \\
      0 &  r_{i,j}>2^{\frac{1}{6}}\sigma
    \end{cases}
\end{equation} 
The diffusion coefficient of all the particles in the solution is $D_a$. A single PBP with diffusion coefficient $D_b<D_a$ is immersed into the solution. This SP has the same size as the other particles in the solution, and it interacts with other particles via similar  WCA potential. 

All the particles are positioned on a $2$-dimensional box of size $S=20\sigma\times 20\sigma$ and periodic boundary conditions are implemented. 
For all particles, Eq.~\eqref{eq:langevinDef} is advanced in time by the Euler-Maruyama method while we use $\epsilon=0.01$, $\sigma=1$ and the discrete-time step $\delta t=0.005$.
The density of the solution, $\phi$, is determined by the total number of particles with diffusion coefficient $D_a$ in the solution, $N_a$, 
\begin{equation}
    \label{eq:phidef}
    \phi=N_a\pi r_0^2/S
\end{equation}
where $2\times r_0=2^\frac{1}{6}\sigma$ is the minimal distance at which two particles start to repel each other. During the simulation, the position of the SP with diffusion coefficient $D_b$ is recorded.

\section{Mean Squared Displacement}
\label{secmsd}

\begin{figure}
\includegraphics[width=0.99\linewidth]{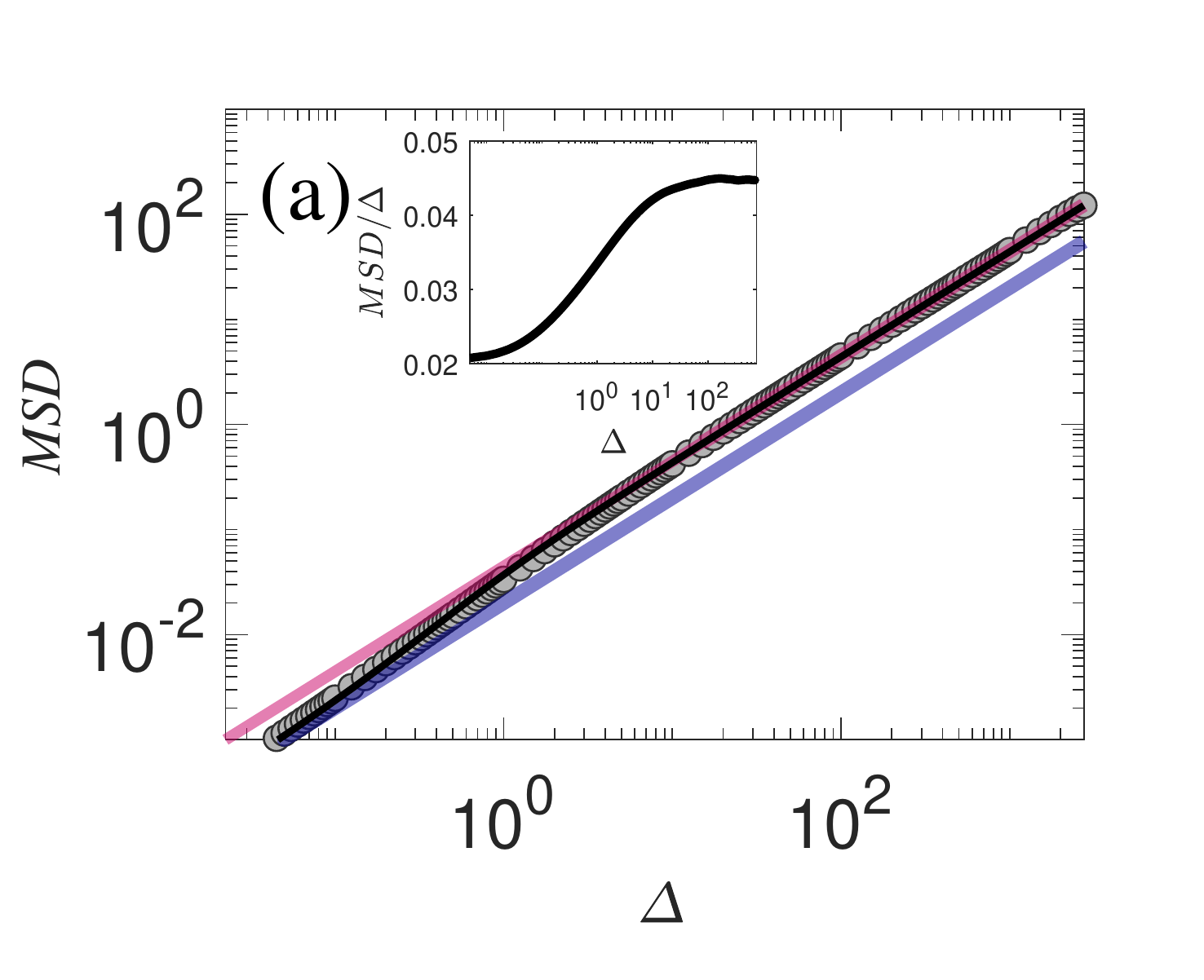}
 \\

    \includegraphics[width=0.89\linewidth]{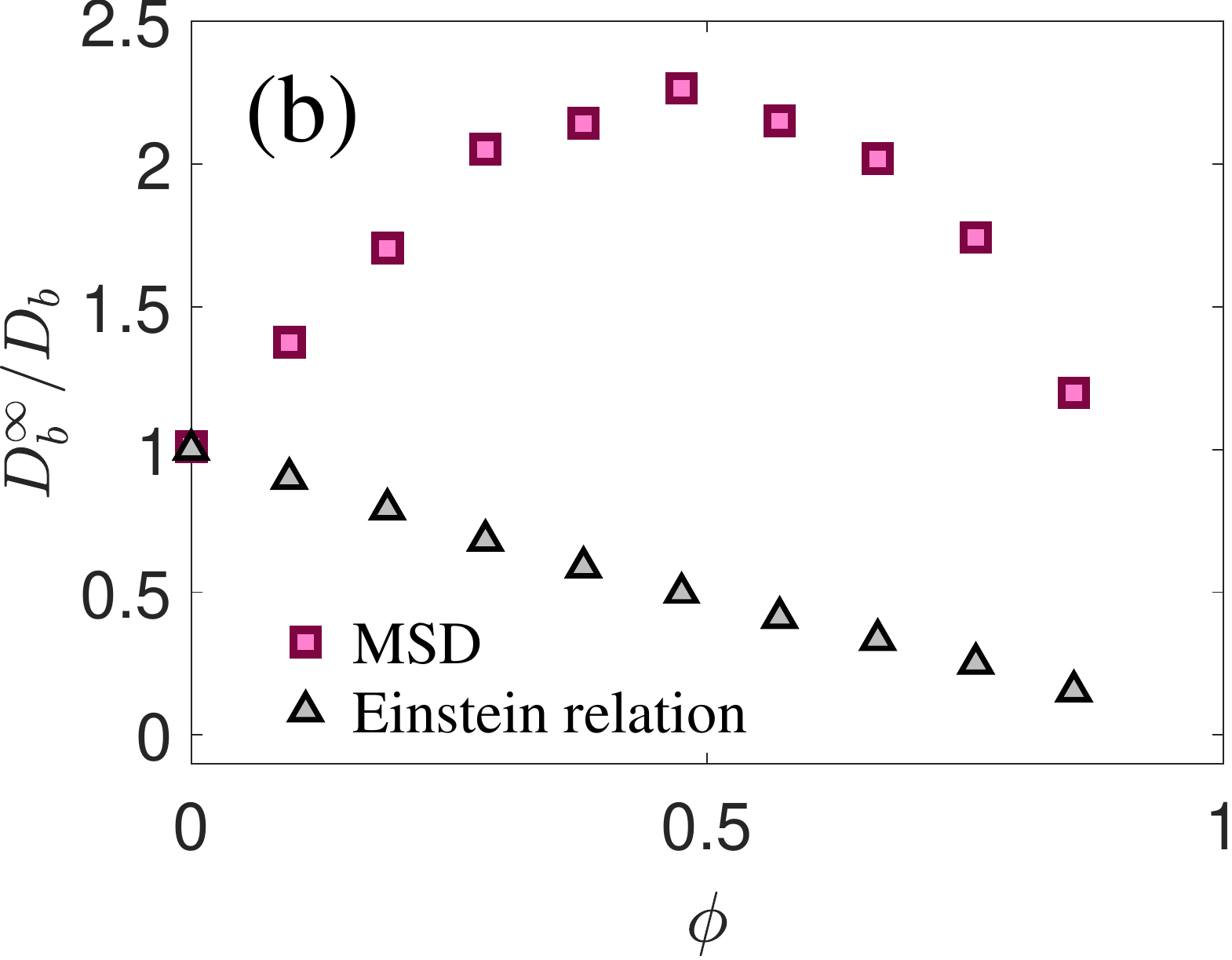}
    \vspace{-4mm}
 \caption{ {\textbf{(a)} MSD of a SP with self-diffusion coefficient $D_b=0.01$ in a solution of PBPs with $D_a=0.1$, while the density $\phi=0.47$. Gray circles represent simulated data, blue line  represents short period diffusion - $2D_b\Delta$. Pink line represent long period diffusion $2D_b^{\infty}$, where the log time diffusion coefficient $D_b^{\infty}=0.022$. 
 Black line represent the fit to the MSD of an active particle, Eq.~\eqref{eq:fitmsd}, where  $A=0.024$ and $B=0.3$.
 Inset: MSD/$\Delta$ with parameters of  panel (a) and semi-logarithmic scale. 
 \textbf{(b)}  $D_b^{\infty}/D_b$ as a function of the density of the solution. The squares represent the calculation of $D_b^{\infty}$ via the MSD of a free particle (Eq.~\eqref{eq:tamsd}).
 The triangles represent situation when external force $F$ is applied upon the SP with $D_b$. $D_b^{\infty}$ is calculated via the Einstein relation  $D_b^{\infty}=v_d k_BT/F$.  $v_d$ is the measured terminal velocity of the particle and $k_BT\propto D_b$.
 For both panels $6\times 10^5$ simulation steps were performed.  Averaging  over time and $40$ ensembles was performed.
 }}
  \label{fig:msd} 
\end{figure}

The first characteristic that we explore is the time and ensemble-averaged mean squared displacement (MSD), of the tracked SP (with $D_b$) during time-frame $\Delta$, 
\begin{equation}
    \label{eq:tamsd}
 \text{MSD}\left(\Delta\right)   = \frac{1}{t-\Delta}\int_0^{t-\Delta}
    \langle   \left(
    \vec{r}_b(t'+\Delta)-\vec{r}_b(t')
    \right)^2\rangle
   \,dt'
\end{equation}
where $\vec{r}_b(t')$ is the position of the tracked particle at time $t'$ and $t$ is the measurement time. 
Fig.~\ref{fig:msd} {\bf (a)} displays 
$\text{MSD}\left(\Delta\right)$ 
 for the case of $D_b/D_a=0.1$.
The MSD grows linearly for short times as $2D_b\Delta$ and experiences a transition to a diffusive behavior with a higher long-time diffusion coefficient at longer times, i.e., $\text{MSD}\left(\Delta\right)\sim 2D_b^\infty\Delta$, where $D_b^\infty\big/D_b=2.2$ for the particular case of  Fig.~\ref{fig:msd} {\bf (a)}. 
The MSD enhancement is perfectly fitted by a phenomenological formula for the MSD of an active particle. ~\cite{Golestanian2007,RevModPhys2016}
\begin{equation}
    \label{eq:fitmsd}
\text{MSD}(\Delta)=\left[2D_b+A\right]\Delta+AB\left[e^{-\Delta/B}-1\right],
\end{equation}
 where $A$ and $B$ are constants (see Supplementary Material (S.M.) for further details). 

\begin{figure*}[t]
\includegraphics[width=0.33\linewidth]{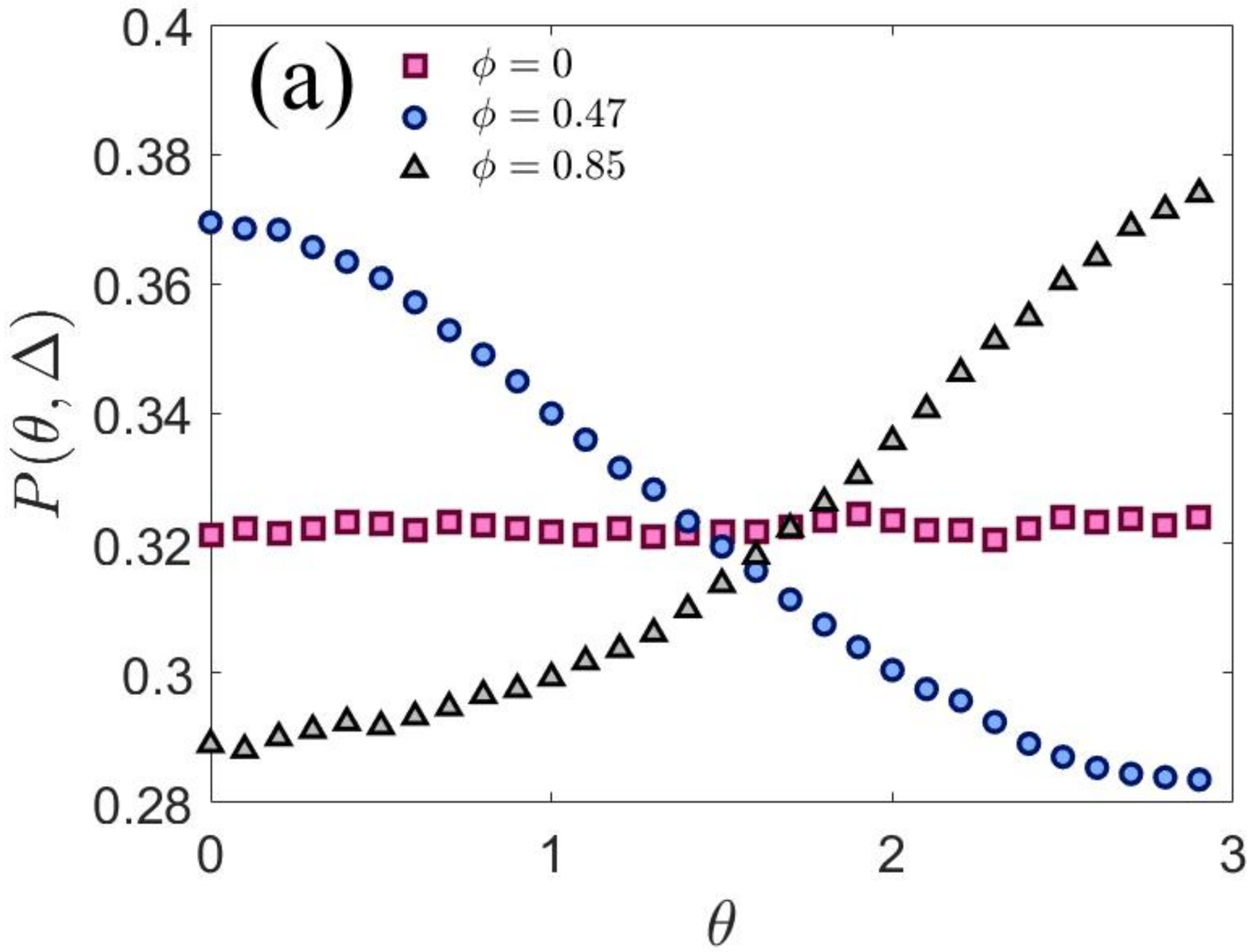}
\includegraphics[width=0.31\linewidth]{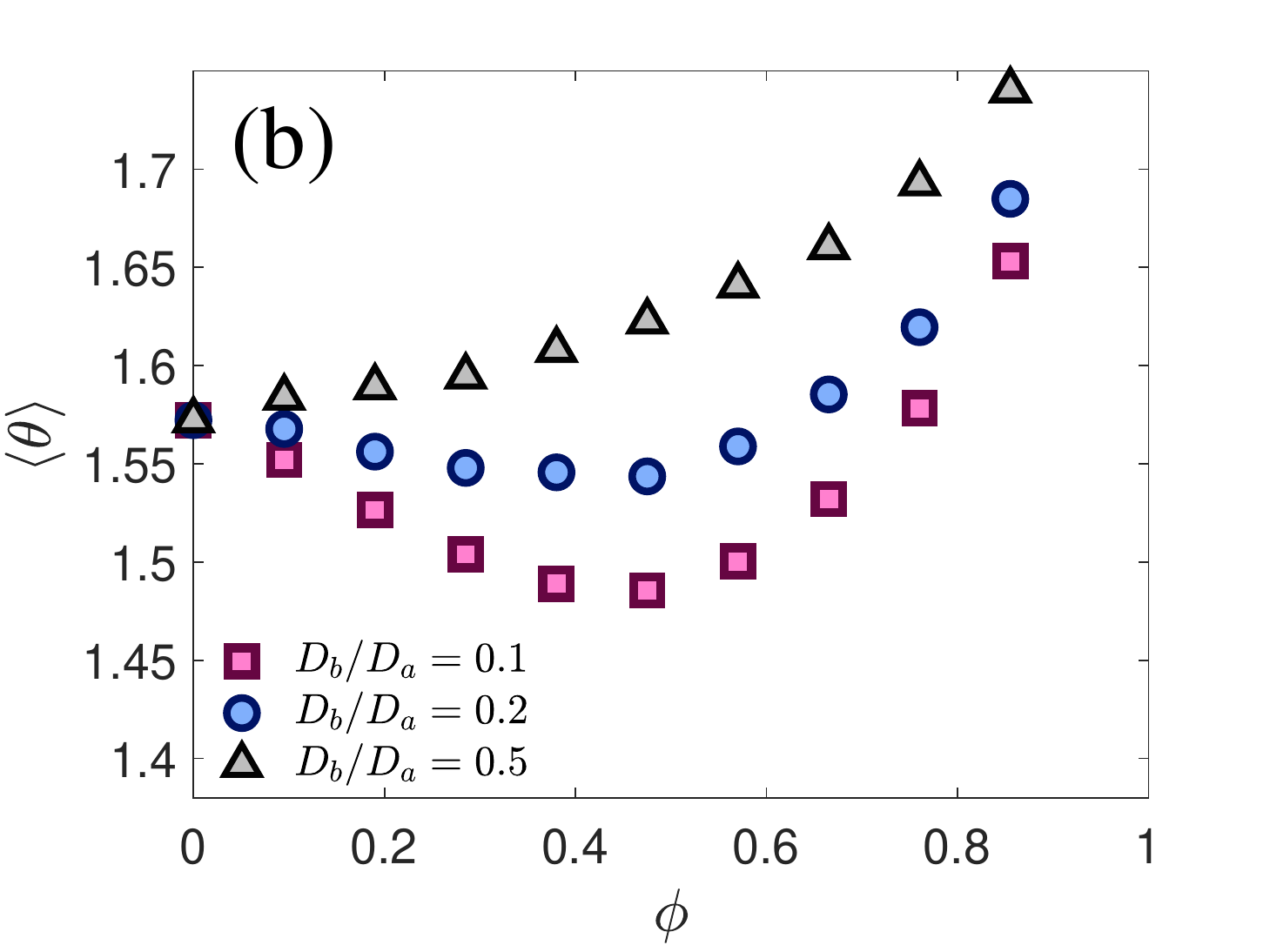}
\includegraphics[width=0.33\linewidth]{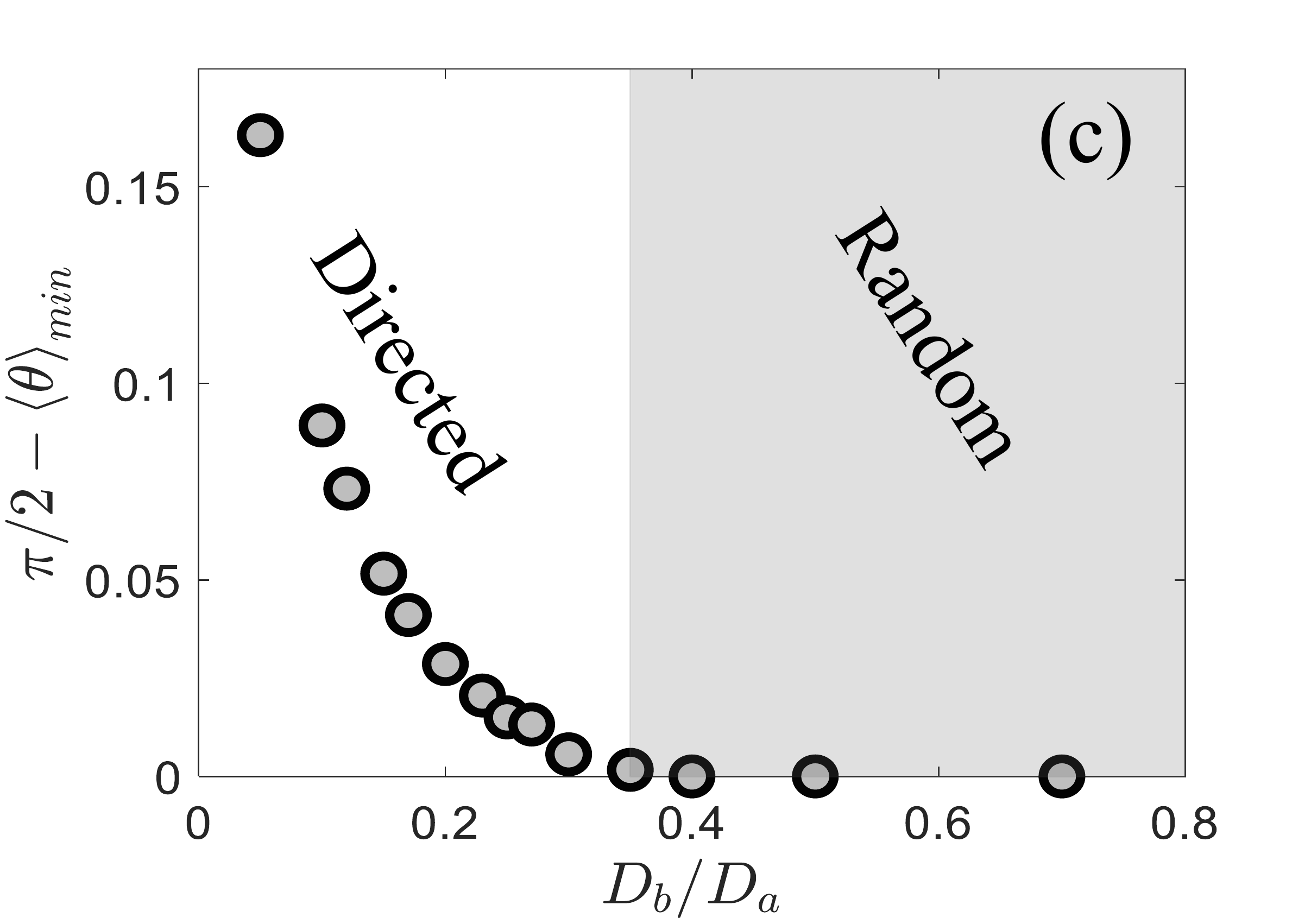}
     \vspace{-0mm}
\caption{ {\bf(a)} Relative angle distribution $P\left(\theta;\Delta\right)$ for the SP with $D=D_b$. Three different solution densities $\phi$ are presented,  $\Delta=200 \delta t$,  $D_{a}=0.1$ and $D_{b}=0.01$.  
{\bf(b)} Mean relative angle, $\langle \theta \rangle$ of the tracked SP as a function of $\phi$ for three different ratios of $D_b/D_a$ ($\Delta=200 \delta t$). Values of $\langle \theta \rangle<\pi/2$ correspond to directed motion. 
 {\bf (c)} Phase diagram of directed vs. uncorrelated motion of the tracked SP. For  every $D_b/D_a$ the minimal value of $\langle \theta \rangle$ ($\langle \theta \rangle_{min}$) was found (searched for different $\Delta$s and $\phi$). $\pi/2-\langle\theta\rangle_{min}=0$ states that the behavior is  passive, when $\pi/2-\langle\theta\rangle_{min}>0$ directed motion is observed. 
 $D_a$ was always set to $0.1$.
For all panels, $6\times 10^5$ simulation steps  were recorded, and averaging over $50$  ensembles was performed.
}
\label{fig:theta}
\end{figure*}

 The increase of diffusion of a ``cold" particle (low diffusion coefficient) moving between ``hot" particles (high diffusion coefficient) was previously observed \cite{Joanny,soft_particles_different_thermostats}. In these previous studies, a system of hard spheres \cite{Joanny} and a system of soft particles \cite{soft_particles_different_thermostats} were used. In both studies, \cite{Joanny,soft_particles_different_thermostats} the diffusion was enhanced when the ratio of the diffusivity of slow particles and the diffusivity of fast particles was below $\sim$ 0.4.
 Our results (Fig.~\ref{fig:msd}{\bf (b)} support these findings for sufficiently low $ֿ\phi$. Several questions are in place. What is the microscopical mechanism that leads to such enhancement of effective diffusion coefficient? Does the observed enhancement of the diffusion coefficient is also associated with directional motion? 
 While the increase in MSD can occur due to a series of large and uncorrelated "kicks"/bombardments~\cite{Mario2021} (see S.M. for additional details), it is also possible that the microscopic mechanism leads to a directed, active-like motion. Finally, what happens to the effect when the density of the suspension is further increased? We address these questions below.       
 
When the long time effective coefficient $D_b^\infty$ is measured for different values of the suspension density $\phi$, a non-monotonic dependence is observed (Fig.~\ref{fig:msd} {\bf (b)}). 
$D_b^\infty$ increases with $\phi$ up to $\phi\approx 0.45$, i.e.,  crowded is faster.  
 Similar behavior for different ratios of diffusion coefficients is presented in Fig.S1{\bf(a)} in the S.M..
Recently, diffusion enhancement via crowding was observed for  an active system of strongly interacting stiff self-propelled filaments~\cite{Lowen2020} and enzyme diffusion driven by chemical reactions~\cite{Golestenian2018,Granick2020A,Granick2020B,Gilson2020}. 
For PBP, $D_b^\infty$ will stop increasing for sufficiently high solution density. 
The intuitive reduction of $D_b^\infty$ as a function of density appears when $\phi\gtrsim 0.45$. 
Such non-monotonous behavior of the diffusion coefficient  was observed in  glassy systems~\cite{Stanely200,Stillinger2001,Zamponi2014,Angell2018}. 
The cause of the eventual decrease of the long-time diffusion coefficient, and therefore the overall non-monotonic behavior,  is the cage effect: an obstruction of motion due to nearby particles that lead to the appearance of negative correlation of displacements ~\cite{uncoraleted_Heuer}.

In addition to the described extraction of the diffusion coefficient by measurement of the MSD of unperturbed particles, we exploited an additional method. A small external force $F$ was applied on the SP and long-time drift velocity $v_d$ was measured. We used this $v_d$ and the Einstein relation $D = v_dk_BT/F$ (by associating $k_BT$ with $D_b$) to extract the effective long-time diffusion coefficient $D_b^\infty$.
The drift velocity was found to monotonically decrease with growing $\phi$, as is intuitively expected for a growing crowdedness of the system.
Fig.~\ref{fig:msd} {\bf (b)} shows that the effective long-time diffusion coefficient $D_b^\infty$ obtained via Einstein relation monotonically decreases with $\phi$. 
The inconsistent behavior of $D_b^\infty$, when measured via MSD or the Einstein relation, shows that the Einstein relation is broken.
Such failure of the Einstein relation is a signature of the non-equilibrium state of our system, see also  \cite{Cugliandolo_effective_temp}.


\section{Relative Angle}
\label{secrelativeangle}

To answer the question  whether the enhancement is accompanied by directed motion, we search for directional properties in the motion of the PBP with $D=D_b$ 
and use the correlations between successive displacements~\cite{L_onard_2005, uncoraleted_Heuer,Hetero_homo_dynamics_Heuer}. Specifically, we use the distribution of directional change by employing the relative angle $\theta(t;\Delta)$~\cite{Burov2013}, previously utilized for analysis of myosin dynamics~\cite{Burov2013}, Lagrangian trajectories in turbulence~\cite{Schneider2015}, swarming bacteria~\cite{beer2015}, football players~\cite{Schneider2017}, and active Brownian particles~\cite{Huang2022}. $\theta$ is the angle between two successive steps of time span $\Delta$ performed by the tracked SP:
\begin{equation}
    \label{relativeangle}
    \cos\theta(t;\Delta) = \frac{
    \vec{\bf{v}}_b(t;\Delta)\cdot \vec{\bf{v}}_b(t+\Delta;\Delta)}{|\vec{\bf{v}}_b(t;\Delta)||\vec{\bf{v}}_b(t+\Delta;\Delta)|}
\end{equation}
where $\vec{\bf{v}}_b(t;\Delta)=\vec{r}_b(t+\Delta)-\vec{r}_b(t)$. For a given $\Delta$, $\theta(t;\Delta)$ is computed for all different $t$s in a  trajectory.
$P(\theta;\Delta)$ is the distribution of directional change  obtained for all   $0\leq\theta\leq\pi$. 
The $P(\theta;\Delta)$ of a non-interacting PBP attains a uniform shape  (Fig.~\ref{fig:theta} {\bf{(a)}}), for any  lag times $\Delta$, since every step is independent of the other steps. 
There is a preference for the retraction of steps when the particle is caged. In such case $P(\theta;\Delta)$ will have a peek at $\theta=\pi$ (Fig.~\ref{fig:theta} {\bf{(a)}}).
When there is a directed motion,   $P\left(\theta;\Delta\right)$ will show a peak at $\theta=0$ for $\Delta$s that corresponds to timescales for which directionality exists~\cite{Burov2013}. 
The peak at $\theta=0$ means that if the particle moved for a lag time $\Delta$ in the direction of $\vec a$, for the next $\Delta$, the preferred direction of motion  will be again $\vec a$. 
In Fig.~\ref{fig:theta} {\bf (a)} the $P(\theta;\Delta)$ is plotted for three different values of the density $\phi$ of the solution. When $\phi=0$, 
$P(\theta;\Delta)$ is uniform as is expected for a non-interacting PBP. 
When $\phi$ is sufficiently large, a peak at $\theta=\pi$ is observed, as is expected for a particle caged by its neighbors. 
Finally, there is a regime when the solution is not very dense, and $P(\theta;\Delta)$ peaks at $\theta=0$. 
For example when $\phi=0.47$ (see Fig.~\ref{fig:theta} {\bf (a)}). 
The PBP hasn't simply performed bigger jumps that increased the MSD; it moved in a directed style for a period $2\Delta$. 

\begin{figure*}[ht!]
\includegraphics[width=0.39\textwidth]{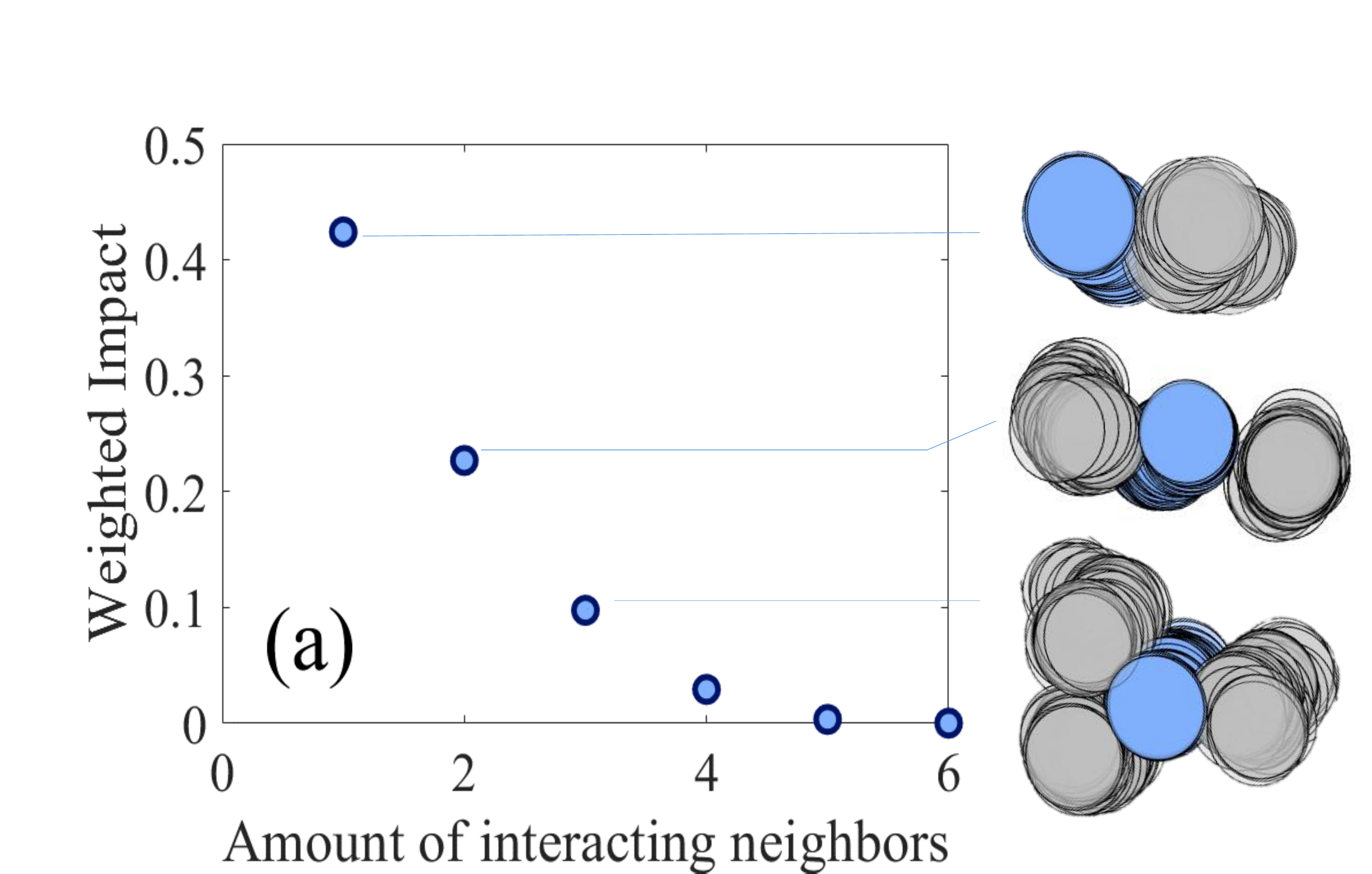} \hspace{0.2em}
\includegraphics[width=0.283\textwidth]{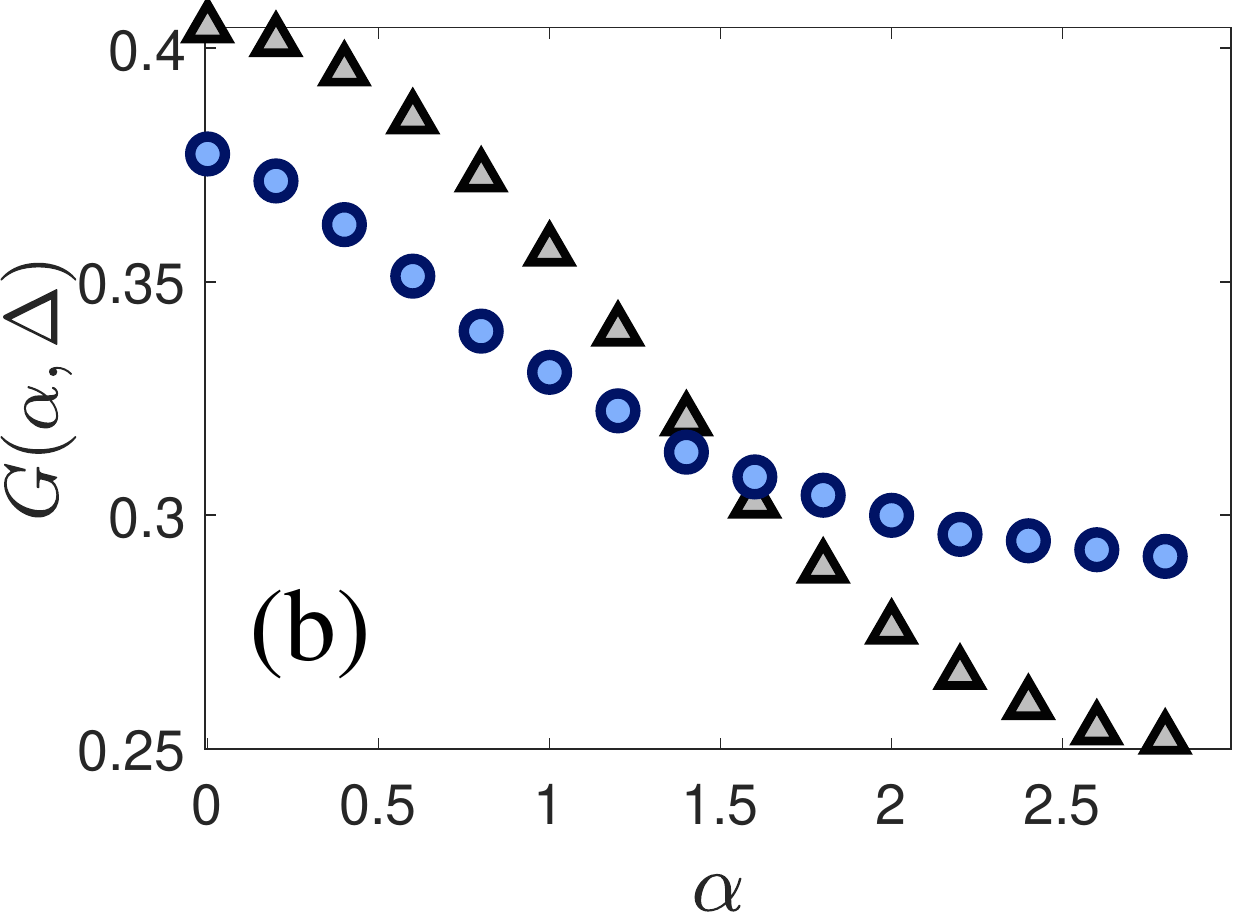} \hspace{1em}
    \includegraphics[width=0.27\textwidth]{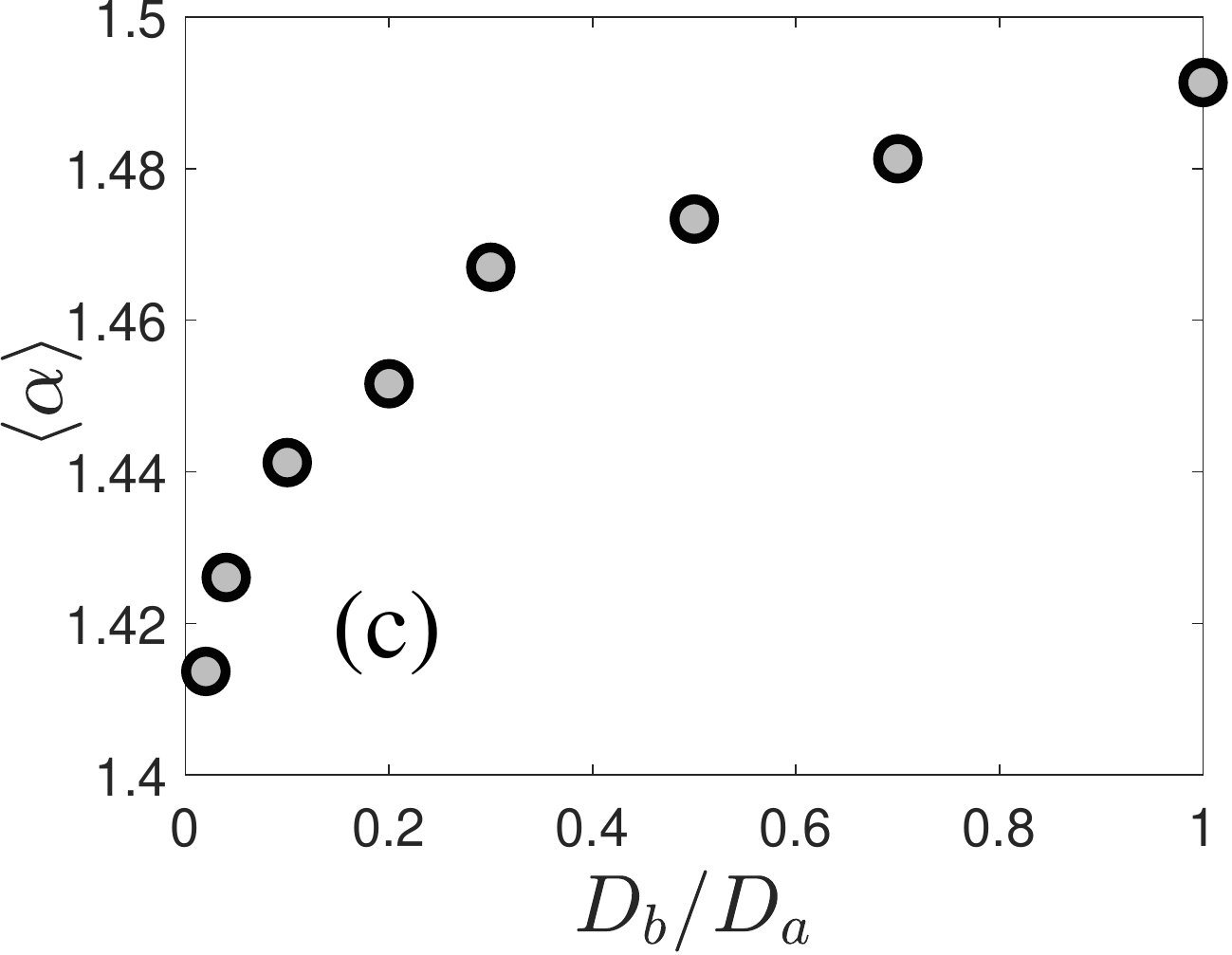}
  \caption{ {
   {\bf(a)} Weighted Impact: The portion of temporal span $\Delta$ during which the tracked SP simultaneously interacted with $n$ particles, as a function of $n=1,2,3,4,5,6$.  $D_a=0.1$,  $D_b=0.01$,  $\phi=0.475$ and $\Delta=200\delta t$. 
   {\bf(b)} $G(\alpha,\Delta)$ :  
   The probability density function of the angle $\alpha(t;\Delta)$ (Eq.~\eqref{eq:alphadef}) between the direction traced by the tracked SP during $\Delta$ and the direction traced by a particle with which the tracked SP mostly interacted during $\Delta$. Such two particles form a temporally correlated pair.  $\Delta=200\delta t$ and two ratios of $D_b/D_a=1$ ($\bigcirc$), $D_b/D_a=0.2$ ($\bigtriangleup$) are presented.
   {\bf(c)}
   Average $\alpha(t;\Delta)$ ($\langle \alpha \rangle$) as obtained from $G(\alpha,\Delta)$ in panel {\bf(b)}.
  $6\times 10^5$ simulation steps  and averaging over $40$ ensembles was performed.
  }}
  \label{fig:spcoupling} 
\end{figure*}

The behavior of the relative angle also reveals that directional motion emerges only if the ratio $D_b/D_a$ is sufficiently smaller than $1$. 
The behavior of  $\langle \theta(\Delta) \rangle = \int_0^\pi\theta P(\theta;\Delta)\,d\theta$ can be utilized as a measure of the directional motion. When the motion is uncorrelated and non-directional, the distribution of $\theta$ is uniform (Fig.~\ref{fig:spcoupling}{\bf(a)}) therefore $\langle \theta \rangle=\pi/2$.  
In Fig.~\ref{fig:theta} {\bf (b)} we present $\langle \theta \rangle$ as a function of $\phi$  for three different values of $D_b/D_a$. 
When  $\langle \theta \rangle$ attains values below $\pi/2$ it means that $P(\theta;\Delta)$ is peaked around $\theta=0$ and the particle moves directionally  for a time frame of $2\Delta$. 
Alternatively, when $\langle \theta \rangle>\pi/2$,  $P(\theta;\Delta)$ is peaked around $\theta=\pi$, the particle is caged. 
Therefore, directed motion emerges when the minimum measured $\langle\theta\rangle$, i.e., $\langle\theta\rangle_{min}$, is smaller than $\pi/2$. Fig.~\ref{fig:theta} {\bf (b)} displays a monotonic growth of $\langle\theta\rangle$ as a function of  $\phi$ when $D_b/D_a=0.5$. Such behavior is what is naively expected from a system that becomes more and more crowded, SP becomes increasingly caged~\cite{uncoraleted_Heuer}. 
On the other hand for small enough $D_b/D_a$, $\langle \theta\rangle$ behaves non-monotonically as a function of $\phi$, with minimum values below $\pi/2$.
We use $\pi/2-\langle \theta \rangle_{min}$ as an order parameter for the directed motion of the tracked SP. 
For a given $D_b/D_a$, $\langle \theta \rangle_{min}$ is calculated for all different $\Delta$s and $\phi$s. The results are displayed in Fig.~\ref{fig:theta} {\bf (c)}.
When $D_b/D_a$ is above $\approx 0.3$ , $\pi/2-\langle \theta \rangle_{min}$ was always $0$ and behaved similarly to the case of $D_b/D_a=0.5$ in Fig.~\ref{fig:theta}{\bf (b)}, irrespective of the size of $\Delta$. We conclude that no directed motion is detected when $D_b/D_a>0.3$. For  $D_b/D_a<0.3$, a minimum value of $\langle \theta \rangle_{min}$ was always found to be below $\pi/2$ and $\langle \theta \rangle$ behaves similarly to the cases $D_b/D_a=0.2$ and $D_b/D_a=0.1$ in Fig.~\ref{fig:theta}{\bf (b)}.  
Therefore for $D_b/D_a<0.3$ there is always a time-span $\Delta$ for which the PBP moves in a directed fashion.  In Fig.S1 of the SM we show that the  long-time diffusion coefficient $D_b^\infty$ is also enhanced when $D_b/D_a<0.3$.  
Therefore a critical ratio of diffusivities is needed  in order to obtain directed motion which gives rise to the enhancement of the diffusion coefficient. 
This enhancement of the diffusion coefficient of a ``cold" particle in a bath of hot particles was previously observed \cite{Joanny,soft_particles_different_thermostats}. 
The presented analysis of relative angle reveals the reason for the enhancement: at intermediate time scales during which the "cold" particle moves persistently and experiences a memory of the traced direction. At sufficiently long time scales this memory is lost and the effective displacements become uncorrelated. 
But since on intermediate time scales the motion was directed, i.e. positively correlated, these effective uncorrelated displacements are larger, in size, as compared to displacements on smaller time scales. 
This increase in the size of a single uncorrelated displacement leads to the observed enhancement of the long-time diffusion coefficient. 
Therefore we can state that the observed directed motion on intermediate time scales leads to the enhancement of the diffusion coefficient on long time scales. 
In the next section, we address the mechanism that is responsible for the observed spontaneous directional symmetry breaking, i.e., directional motion. 

%

\section{Temporally Correlated Pairs}
\label{sectiontcp}

To elucidate the origin of directional motion, we measure the fraction of time that the traced SP spent simultaneously interacting with $1$,$2$,$3$,.. neighbors. The Weighted Impact  is the portion of time $\Delta$ that the SP simultaneously interacted (i.e., experienced non-zero force) with a  number of other PBPs. 
Panel {\bf (a)} of Fig.~\ref{fig:spcoupling} shows that the majority of the interactions are two-body encounters. 
We focus on two-particle encounters 
and define the angle $\alpha$  
\begin{equation}
\cos\alpha(t;\Delta)= \frac{\vec{\bf{v}}_b(t;\Delta)\cdot \vec{\bf{v}}_a(t+\Delta;\Delta)}{{|\vec{\bf{v}}_b(t;\Delta)||\vec{\bf{v}}_a(t+\Delta;\Delta)|}}
\label{eq:alphadef}
\end{equation}
i.e., 
the angle between the displacement of the tracked SP during $\Delta$ and the displacement of a neighbor  SP with which it mostly interacted during $\Delta$. 
We tally the  interactions between the tracked SP and each of its neighbors and then compute the value of $\alpha$ while using the neighbor SP with whom the tracked SP had the highest number of 
interactions during $\Delta$. 
When particles do not interact with each other, the probability density function of $G(\alpha,\Delta)$ is uniform, as one would expect for a typical PBP. 
However, if there is a peak in the probability density function of $G(\alpha,\Delta)$ around $\alpha=0$, it suggests that the particles tend to move in the same direction. 
On the other hand, if the peak is at $\alpha=\pi/2$, this indicates that the particles tend to move in opposite directions.
 Fig.~\ref{fig:spcoupling} {\bf (b)} displays the probability density function of $\alpha$, $G(\alpha,\Delta)$, computed for all possible displacements of the tracked SP along the trajectory. 
Even when $D_b=D_a$, the distribution $G(\alpha,\Delta)$ is peaked at $\alpha=0$. Meaning that there are time scales when two neighbors prefer to move in the same direction (on average).
This effect is achieved due to the presence of other PBPs in the vicinity of the pair. These PBPs do not let particles in the pair  separate and create  effective temporal coupling for the  pair. 
We term such a pair a temporally correlated pair (TCP).
When we decrease the fraction $D_b/D_a$, this effect of correlated motion of neighbor particles is enhanced. 
Panel {\bf(c)} of Fig.~\ref{fig:spcoupling} shows how the average $\alpha$ is decreasing with decreasing $D_b/D_a$. The preference to move in a cohort manner increases when $D_b/D_a\to 0$. 

 \begin{figure*}[t]
\includegraphics[width=0.32\textwidth]{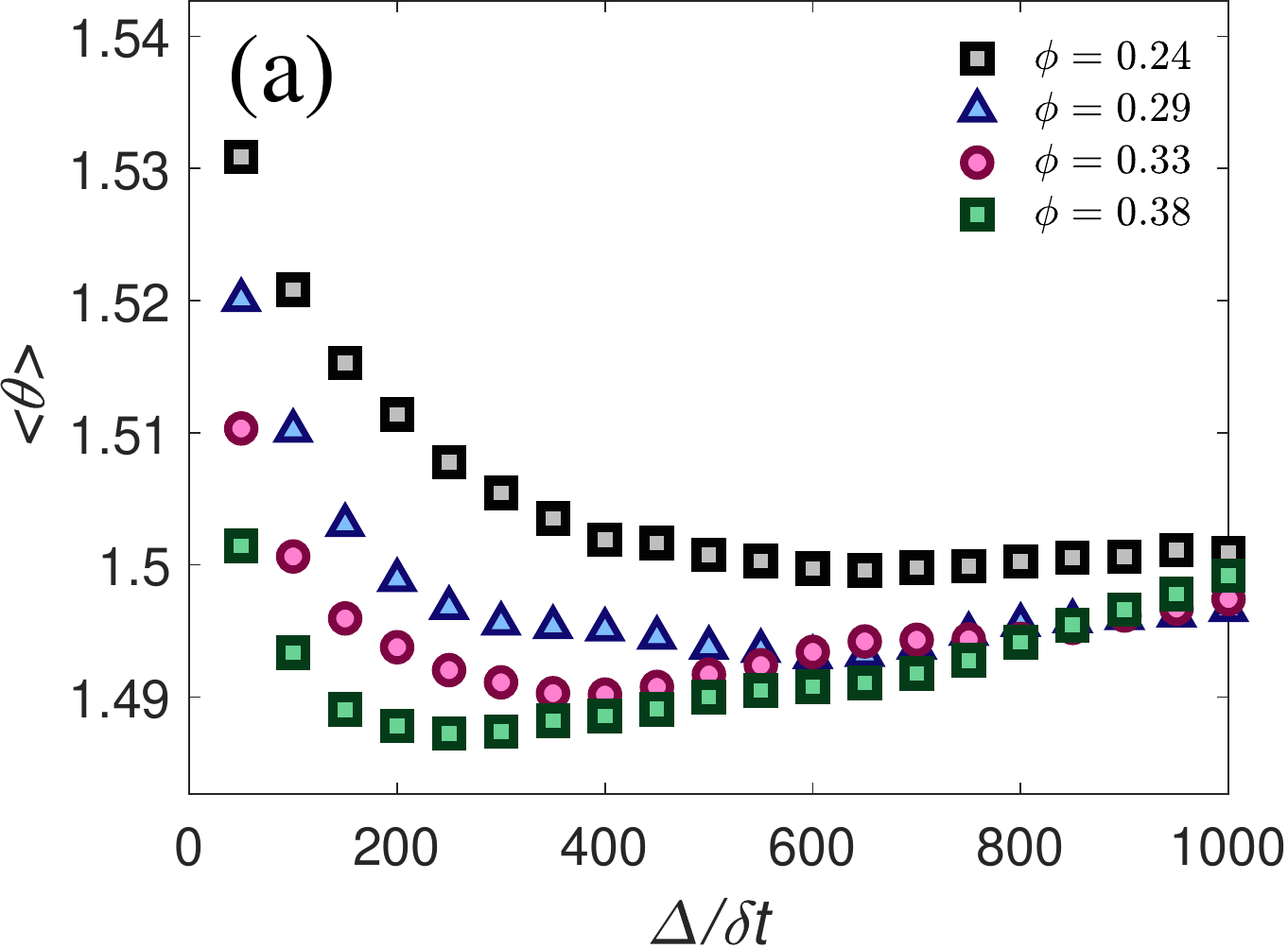} \hspace{1em} 
\includegraphics[width=0.3\textwidth]{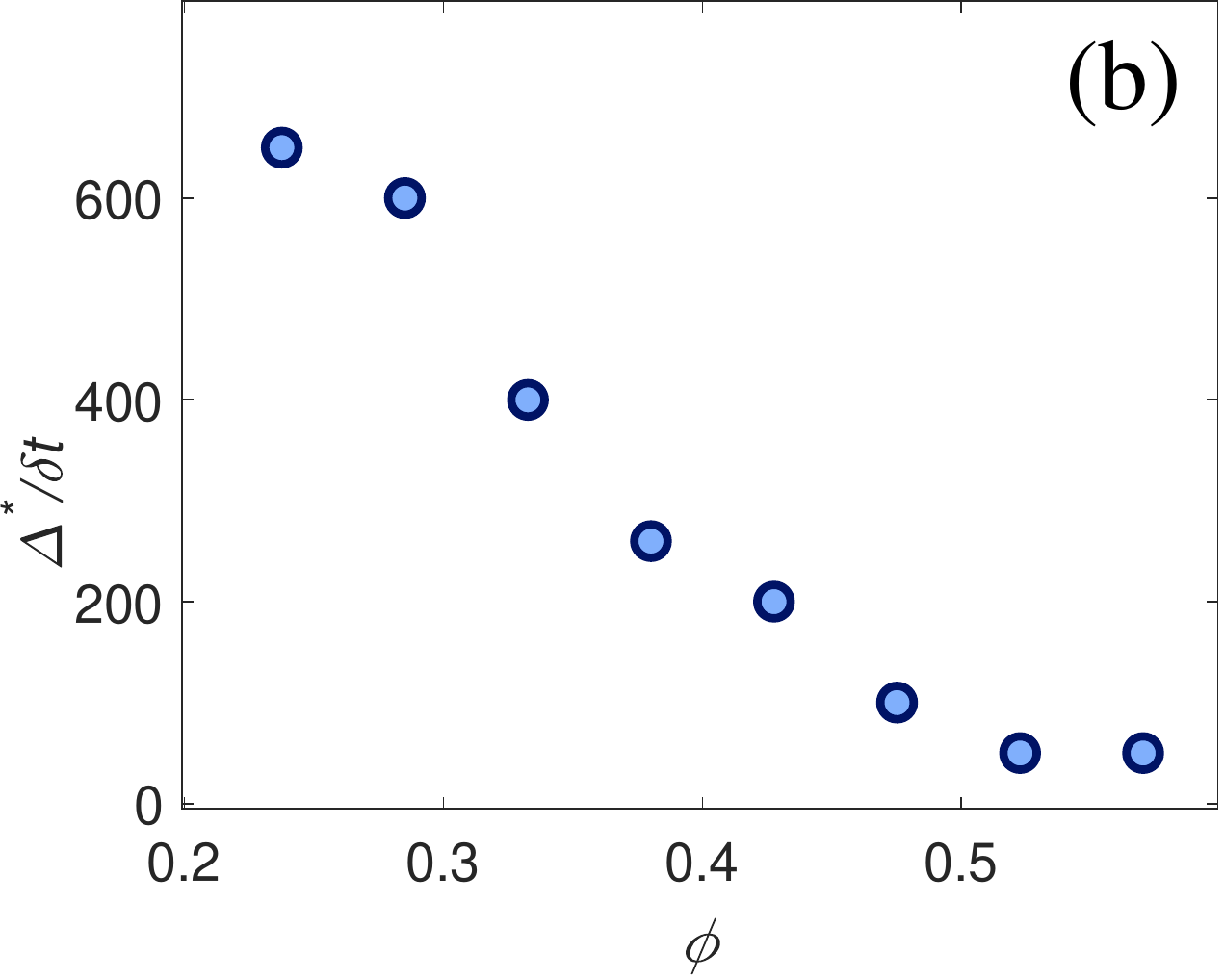} \hspace{1em} 
    \includegraphics[width=0.308\textwidth]{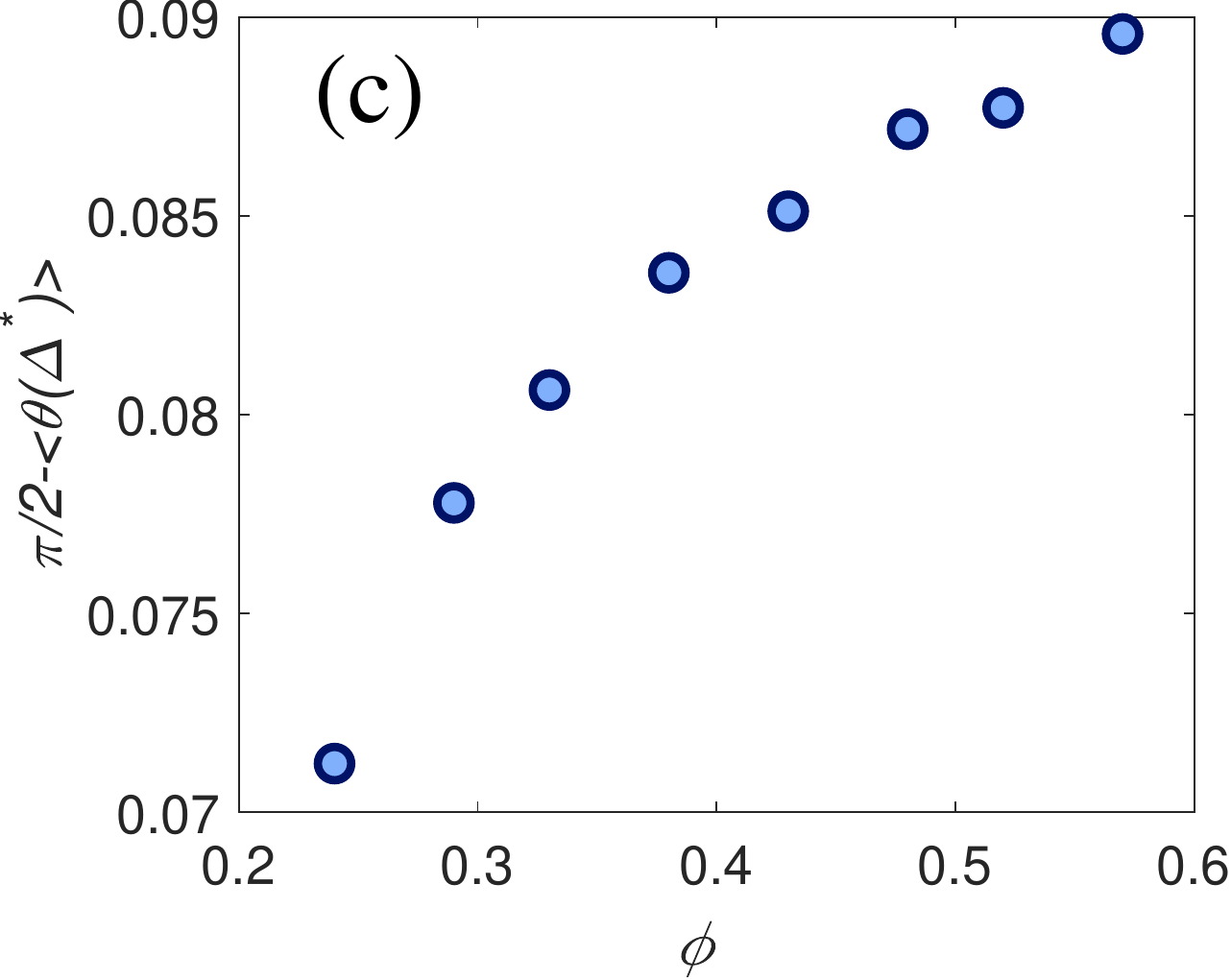} \hspace{1em} 
   \vspace{-0mm}
  \caption{
   {\bf(a)} Non-monotonic evolution of the mean relative angle $<\theta>$ as a function of $\Delta$ for $4$ different densities. 
   {\bf (b)} $\Delta^*$ is the $\Delta$ for which the minimum value of $\langle \theta \rangle$ was obtained (see panel (a) for a given $\phi$ and $D_b/D_a$. Plotted as a function of the suspension density $\phi$.
    {\bf(c)} Order parameter $\pi/2-\langle \theta \rangle$, when the mean relative angle $\langle \theta\rangle$ is computed at $\Delta=\Delta^*$ ( panel (b)). Higher values of the order parameter describe more pronounced directed motion. 
     For all panels, $D_a=0.1$ and $D_b=0.01$, $6\times 10^5$ simulation steps  and averaging over $100$ ensembles was performed. }
  \label{Time_window_of_RA} 
\end{figure*}

This phenomenon of TCP creation is what leads to spontaneous directional symmetry breaking. 
On intermediate time scales, the preferred direction of motion becomes the line of  interaction between the constituents of the TCP pair.
In Fig.~\ref{Time_window_of_RA}{\bf(a)} we plot the temporal evolution of the average relative angle $\langle \theta \rangle$ as a function of $\Delta$. The lower the value of $\langle \theta \rangle$, the greater the directional preference of the motion. $\langle \theta \rangle$ behaves non-monotonically with $\Delta$. 
It decreases for short and intermediate time scales and grows when $\Delta$ is sufficiently large. Since the interaction between the different SPs is purely repulsive, it takes time for the effective interaction, which involves the surrounding particles in the suspension, to build up and create a TCP. Therefore the decrease of $\langle \theta \rangle$ on short time scales. A given TCP can't survive for a very long time since the role of the partner with which a given SP interacts the most will switch from one particle to another. This leads to the observed growth of $\langle \theta \rangle$ for large $\Delta$. We call the $\Delta$ for which $\langle \theta \rangle$ obtains its minimal value, $\Delta^*$, and associate it with the average lifetime of a TCP. $\Delta^*$ is the time scale on which the SP moves most persistently in a specific direction defined by the orientation of the TCP. 
For $\Delta$s smaller/larger than $\Delta^*$ $\langle \theta\rangle$ is closer to $\pi/2$, meaning the motion is less directional, which occurs due to switching of most interacting partners and breaking of the TCP. 
We expect two different behaviors of $\Delta^*$ and $\langle \theta(\Delta^*) \rangle$ as a function of the density of the suspension $\phi$. 
(I) The denser the suspension is, the faster the effective interaction will build up and cause the faster formation of TCPs. But, since the suspension is denser, it will also be expected to cause a faster dissociation of partners and dissociation of TCps. A decrease of $\Delta^*$ with $\phi$ is expected.
(II) The denser the suspension is, the closer the different SPs are; therefore, the created TCP will be tighter. Meaning, the SPs have less free space, and the directional motion will be more pronounced. This will show itself in a smaller $\langle \theta(\Delta^*)\rangle$. An increase of $\pi/2-\langle \theta(\Delta^*)\rangle$ is expected with $\phi$, i.e. the peak around $\theta=0$ of $P(\theta,\Delta^*)$ grows with $\phi$. 
Both (I) and (II) are observed in Fig.~\ref{Time_window_of_RA} {\bf(b)} and {\bf(c)}.

The effect of the surrounding media is thus two-fold. The presence of surrounding particles provides the means for directional symmetry breaking by temporal coupling of nearby particles. But exactly the same surrounding particles randomize this coupling, thus destroying the formed TCP and restoring the directional symmetry on longer time scales. A sufficiently long time $\Delta^*$ between the reorientation events is needed to observe the effect of directional motion.

\section{Discussion}
\label{secdiscussion}

The results presented in this work show that directional motion on the SP level can emerge simply due to differences in such scalar quantity as the diffusion coefficient. 
The motion of a single PBP, surrounded by other PBPs with a higher diffusion coefficient, becomes directed. 
When the density is sufficiently low, the SP with a lower diffusion coefficient $D_b$ temporally couples to one SP with diffusion coefficient $D_a$ and creates a TCP. 
As $\phi$ grows, the coupling becomes more prominent, the persistence  of directed motion is enhanced on intermediate time-scales, and so is the long-time diffusion coefficient ${D_b^\infty}$. 
But this occurs only if the interactions on the TCP level are not randomized by interactions with other particles in the immediate neighborhood of the TCP. 
When the density $\phi$ is sufficiently high, the caging effect takes place, i.e., the growing peak for $\theta=\pi$ of $P(\theta,\Delta)$ (Fig.~\ref{fig:theta}). 
The rise of caging implies that the tracked SP encounters frequently with many surrounding PBPs. The lifetime of the TCP, or the time while the directional symmetry breaking holds, $\Delta^*$, decreases with density (Fig.~\ref{Time_window_of_RA} {\bf(b)}).
The tracked SP simply switches partners too frequently thus effectively randomizing its motion and preventing persistence in a specific direction.

Thus the picture that emerges from our study is of effective directed motion that is a consequence of TCP creation for sufficiently prolonged time periods. While TCP can occur also when the tracked SP has exactly the same diffusion coefficient as the nearby PBPs, the interaction and the cooperative motion on the TCP level increase as the ratio of the diffusion coefficients $D_b/D_a$ decreases. 
The density of the suspension also facilitates the creation of TCP and the time for which a given TCP survives. 
While the tracked SP is a part of a given TCP it has a preferred direction of motion, dictated by the location of the other part of the TCP. 
Therefore, the survival time of a given TCP defines the time scale over which the motion of the tracked SP is directional. When this time scale is long enough, the effective displacements become large enough and lead to the observed enhancement of the long-time diffusion coefficient $D_b^\infty$. But the effect of the growth of $\phi$ is two-fold. 
While the interaction on the TCP level increases, the encounters with other nearby particles become more frequent (caging effect) and so the lifetime of a given TCP will eventually start to decrease when $\phi$ is high enough. 
This leads to the diminishing of the long-time diffusion coefficient and the overall non-monotonic behavior of $D_b^\infty$ observed in Fig.~\ref{fig:msd}.


The observed breakdown of the Einstein relation (Fig.~\ref{fig:msd} {\bf(b)}) is consistent with the described mechanism of enhancement of diffusion coefficient due to the spontaneous creation of TCPs.
The existence of an external field (applied only upon the SP with $D_b$) and its direction is uncorrelated with the spontaneous direction that emerges due to the creation of TCP. 
The dragged SP already has a preferred direction of motion (i.e., the direction of the external force), and the creation of a TCP  only obstructs the dynamics in this preferred direction.
While for the case when no external force is present, the TCP enforces a preferred direction of motion. When an external force is applied the nearby particles (potential partners of a TCP) only "stand in the way" of the preferred direction of motion, i.e., the direction of the external force. 
Therefore the MSD, obtained via the Einstein relation,  monotonically decays as a function of the density. Unlike the observed non-monotonic behavior of $D_b^\infty$ with $\phi$,  when no external force is present. 


In this work, we don't 
observe a macroscopic motion of many particles.
Instead, we have seen that for the SP, directional motion occurs only when $D_b/D_a \lesssim  0.3$ (Fig.~\ref{fig:theta} {\bf(c)}). 
On the SP level, this emergent spontaneous directional symmetry breaking can be addressed as an SP analog of dissipative structure, and $D_b/D_a$ is the parameter that must be modified in order to enhance TCP creation that facilitates the directed motion of a single PBP. 

{\bf Acknowledgments: }This work was supported by the  Israel Science Foundation Grant No. 2796/20.

\bibliography{./deb_references.bib}

\begin{thebibliography}{49}%
\makeatletter
\providecommand \@ifxundefined [1]{%
 \@ifx{#1\undefined}
}%
\providecommand \@ifnum [1]{%
 \ifnum #1\expandafter \@firstoftwo
 \else \expandafter \@secondoftwo
 \fi
}%
\providecommand \@ifx [1]{%
 \ifx #1\expandafter \@firstoftwo
 \else \expandafter \@secondoftwo
 \fi
}%
\providecommand \natexlab [1]{#1}%
\providecommand \enquote  [1]{``#1''}%
\providecommand \bibnamefont  [1]{#1}%
\providecommand \bibfnamefont [1]{#1}%
\providecommand \citenamefont [1]{#1}%
\providecommand \href@noop [0]{\@secondoftwo}%
\providecommand \href [0]{\begingroup \@sanitize@url \@href}%
\providecommand \@href[1]{\@@startlink{#1}\@@href}%
\providecommand \@@href[1]{\endgroup#1\@@endlink}%
\providecommand \@sanitize@url [0]{\catcode `\\12\catcode `\$12\catcode
  `\&12\catcode `\#12\catcode `\^12\catcode `\_12\catcode `\%12\relax}%
\providecommand \@@startlink[1]{}%
\providecommand \@@endlink[0]{}%
\providecommand \url  [0]{\begingroup\@sanitize@url \@url }%
\providecommand \@url [1]{\endgroup\@href {#1}{\urlprefix }}%
\providecommand \urlprefix  [0]{URL }%
\providecommand \Eprint [0]{\href }%
\providecommand \doibase [0]{http://dx.doi.org/}%
\providecommand \selectlanguage [0]{\@gobble}%
\providecommand \bibinfo  [0]{\@secondoftwo}%
\providecommand \bibfield  [0]{\@secondoftwo}%
\providecommand \translation [1]{[#1]}%
\providecommand \BibitemOpen [0]{}%
\providecommand \bibitemStop [0]{}%
\providecommand \bibitemNoStop [0]{.\EOS\space}%
\providecommand \EOS [0]{\spacefactor3000\relax}%
\providecommand \BibitemShut  [1]{\csname bibitem#1\endcsname}%
\let\auto@bib@innerbib\@empty
\bibitem [{\citenamefont {Ilker}\ \emph {et~al.}(2021)\citenamefont {Ilker},
  \citenamefont {Castellana},\ and\ \citenamefont {Joanny}}]{Joanny}%
  \BibitemOpen
  \bibfield  {author} {\bibinfo {author} {\bibfnamefont {E.}~\bibnamefont
  {Ilker}}, \bibinfo {author} {\bibfnamefont {M.}~\bibnamefont {Castellana}}, \
  and\ \bibinfo {author} {\bibfnamefont {J.-F.}\ \bibnamefont {Joanny}},\
  }\href@noop {} {\bibfield  {journal} {\bibinfo  {journal} {Phys. Rev.
  Research}\ }\textbf {\bibinfo {volume} {3}},\ \bibinfo {pages} {023207}
  (\bibinfo {year} {2021})}\BibitemShut {NoStop}%
\bibitem [{\citenamefont {Bechinger}\ \emph {et~al.}(2016)\citenamefont
  {Bechinger}, \citenamefont {Di~Leonardo}, \citenamefont {L\"owen},
  \citenamefont {Reichhardt}, \citenamefont {Volpe},\ and\ \citenamefont
  {Volpe}}]{RevModPhys2016}%
  \BibitemOpen
  \bibfield  {author} {\bibinfo {author} {\bibfnamefont {C.}~\bibnamefont
  {Bechinger}}, \bibinfo {author} {\bibfnamefont {R.}~\bibnamefont
  {Di~Leonardo}}, \bibinfo {author} {\bibfnamefont {H.}~\bibnamefont
  {L\"owen}}, \bibinfo {author} {\bibfnamefont {C.}~\bibnamefont {Reichhardt}},
  \bibinfo {author} {\bibfnamefont {G.}~\bibnamefont {Volpe}}, \ and\ \bibinfo
  {author} {\bibfnamefont {G.}~\bibnamefont {Volpe}},\ }\href {\doibase
  10.1103/RevModPhys.88.045006} {\bibfield  {journal} {\bibinfo  {journal}
  {Rev. Mod. Phys.}\ }\textbf {\bibinfo {volume} {88}},\ \bibinfo {pages}
  {045006} (\bibinfo {year} {2016})}\BibitemShut {NoStop}%
\bibitem [{\citenamefont {Ramaswamy}(2010)}]{Ramaswamy2010}%
  \BibitemOpen
  \bibfield  {author} {\bibinfo {author} {\bibfnamefont {S.}~\bibnamefont
  {Ramaswamy}},\ }\href@noop {} {\bibfield  {journal} {\bibinfo  {journal}
  {Annu. Rev. Condens. Matter Phys.}\ }\textbf {\bibinfo {volume} {1}},\
  \bibinfo {pages} {323} (\bibinfo {year} {2010})}\BibitemShut {NoStop}%
\bibitem [{\citenamefont {Cates}(2012)}]{Cates2012}%
  \BibitemOpen
  \bibfield  {author} {\bibinfo {author} {\bibfnamefont {M.~E.}\ \bibnamefont
  {Cates}},\ }\href@noop {} {\bibfield  {journal} {\bibinfo  {journal} {Rep.
  Prog. Phys}\ }\textbf {\bibinfo {volume} {75}},\ \bibinfo {pages} {042601}
  (\bibinfo {year} {2012})}\BibitemShut {NoStop}%
\bibitem [{\citenamefont {Szamel}(2014)}]{Self_propelled_particle}%
  \BibitemOpen
  \bibfield  {author} {\bibinfo {author} {\bibfnamefont {G.}~\bibnamefont
  {Szamel}},\ }\href@noop {} {\bibfield  {journal} {\bibinfo  {journal} {Phys.
  Rev. E}\ }\textbf {\bibinfo {volume} {90}},\ \bibinfo {pages} {012111}
  (\bibinfo {year} {2014})}\BibitemShut {NoStop}%
\bibitem [{\citenamefont {Marchetti}\ \emph {et~al.}(2013)\citenamefont
  {Marchetti}, \citenamefont {Joanny}, \citenamefont {Ramaswamy}, \citenamefont
  {Liverpool}, \citenamefont {Prost}, \citenamefont {Rao},\ and\ \citenamefont
  {Simha}}]{Marchetti2013}%
  \BibitemOpen
  \bibfield  {author} {\bibinfo {author} {\bibfnamefont {M.~C.}\ \bibnamefont
  {Marchetti}}, \bibinfo {author} {\bibfnamefont {J.~F.}\ \bibnamefont
  {Joanny}}, \bibinfo {author} {\bibfnamefont {S.}~\bibnamefont {Ramaswamy}},
  \bibinfo {author} {\bibfnamefont {T.~B.}\ \bibnamefont {Liverpool}}, \bibinfo
  {author} {\bibfnamefont {J.}~\bibnamefont {Prost}}, \bibinfo {author}
  {\bibfnamefont {M.}~\bibnamefont {Rao}}, \ and\ \bibinfo {author}
  {\bibfnamefont {R.~A.}\ \bibnamefont {Simha}},\ }\href {\doibase
  10.1103/RevModPhys.85.1143} {\bibfield  {journal} {\bibinfo  {journal} {Rev.
  Mod. Phys.}\ }\textbf {\bibinfo {volume} {85}},\ \bibinfo {pages} {1143}
  (\bibinfo {year} {2013})}\BibitemShut {NoStop}%
\bibitem [{\citenamefont {Sokolov}\ \emph {et~al.}(2010)\citenamefont
  {Sokolov}, \citenamefont {Apodaca}, \citenamefont {Grzybowski},\ and\
  \citenamefont {Aranson}}]{Aranson2010}%
  \BibitemOpen
  \bibfield  {author} {\bibinfo {author} {\bibfnamefont {A.}~\bibnamefont
  {Sokolov}}, \bibinfo {author} {\bibfnamefont {M.~M.}\ \bibnamefont
  {Apodaca}}, \bibinfo {author} {\bibfnamefont {B.~A.}\ \bibnamefont
  {Grzybowski}}, \ and\ \bibinfo {author} {\bibfnamefont {I.~S.}\ \bibnamefont
  {Aranson}},\ }\href@noop {} {\bibfield  {journal} {\bibinfo  {journal} {Proc.
  Nat. Acad. Sci.}\ }\textbf {\bibinfo {volume} {107}},\ \bibinfo {pages} {969}
  (\bibinfo {year} {2010})}\BibitemShut {NoStop}%
\bibitem [{\citenamefont {Di~Leonardo}\ \emph {et~al.}(2010)\citenamefont
  {Di~Leonardo}, \citenamefont {Angelani}, \citenamefont
  {Dell{\textquoteright}Arciprete}, \citenamefont {Ruocco}, \citenamefont
  {Iebba}, \citenamefont {Schippa}, \citenamefont {Conte}, \citenamefont
  {Mecarini}, \citenamefont {De~Angelis},\ and\ \citenamefont
  {Di~Fabrizio}}]{Fabrizio2010}%
  \BibitemOpen
  \bibfield  {author} {\bibinfo {author} {\bibfnamefont {R.}~\bibnamefont
  {Di~Leonardo}}, \bibinfo {author} {\bibfnamefont {L.}~\bibnamefont
  {Angelani}}, \bibinfo {author} {\bibfnamefont {D.}~\bibnamefont
  {Dell{\textquoteright}Arciprete}}, \bibinfo {author} {\bibfnamefont
  {G.}~\bibnamefont {Ruocco}}, \bibinfo {author} {\bibfnamefont
  {V.}~\bibnamefont {Iebba}}, \bibinfo {author} {\bibfnamefont
  {S.}~\bibnamefont {Schippa}}, \bibinfo {author} {\bibfnamefont {M.~P.}\
  \bibnamefont {Conte}}, \bibinfo {author} {\bibfnamefont {F.}~\bibnamefont
  {Mecarini}}, \bibinfo {author} {\bibfnamefont {F.}~\bibnamefont
  {De~Angelis}}, \ and\ \bibinfo {author} {\bibfnamefont {E.}~\bibnamefont
  {Di~Fabrizio}},\ }\href {\doibase 10.1073/pnas.0910426107} {\bibfield
  {journal} {\bibinfo  {journal} {Proc. Nat. Acad. Sci.}\ }\textbf {\bibinfo
  {volume} {107}},\ \bibinfo {pages} {9541} (\bibinfo {year}
  {2010})}\BibitemShut {NoStop}%
\bibitem [{\citenamefont {Blickle}\ and\ \citenamefont
  {Bechinger}(2012)}]{Bechinger2012}%
  \BibitemOpen
  \bibfield  {author} {\bibinfo {author} {\bibfnamefont {V.}~\bibnamefont
  {Blickle}}\ and\ \bibinfo {author} {\bibfnamefont {C.}~\bibnamefont
  {Bechinger}},\ }\href@noop {} {\bibfield  {journal} {\bibinfo  {journal}
  {Nat. Phys.}\ }\textbf {\bibinfo {volume} {8}},\ \bibinfo {pages} {143}
  (\bibinfo {year} {2012})}\BibitemShut {NoStop}%
\bibitem [{\citenamefont {Dechant}\ \emph {et~al.}(2015)\citenamefont
  {Dechant}, \citenamefont {Kiesel},\ and\ \citenamefont {Lutz}}]{Dechant2015}%
  \BibitemOpen
  \bibfield  {author} {\bibinfo {author} {\bibfnamefont {A.}~\bibnamefont
  {Dechant}}, \bibinfo {author} {\bibfnamefont {N.}~\bibnamefont {Kiesel}}, \
  and\ \bibinfo {author} {\bibfnamefont {E.}~\bibnamefont {Lutz}},\ }\href
  {\doibase 10.1103/PhysRevLett.114.183602} {\bibfield  {journal} {\bibinfo
  {journal} {Phys. Rev. Lett.}\ }\textbf {\bibinfo {volume} {114}},\ \bibinfo
  {pages} {183602} (\bibinfo {year} {2015})}\BibitemShut {NoStop}%
\bibitem [{\citenamefont {Mart\'{i}nez}\ \emph {et~al.}(2016)\citenamefont
  {Mart\'{i}nez}, \citenamefont {Rold\'{a}n}, \citenamefont {Dinis},
  \citenamefont {Petrov}, \citenamefont {Parrondo},\ and\ \citenamefont
  {Rica}}]{Rica2016}%
  \BibitemOpen
  \bibfield  {author} {\bibinfo {author} {\bibfnamefont {I.}~\bibnamefont
  {Mart\'{i}nez}}, \bibinfo {author} {\bibfnamefont {E.}~\bibnamefont
  {Rold\'{a}n}}, \bibinfo {author} {\bibfnamefont {L.}~\bibnamefont {Dinis}},
  \bibinfo {author} {\bibfnamefont {D.}~\bibnamefont {Petrov}}, \bibinfo
  {author} {\bibfnamefont {J.~M.~R.}\ \bibnamefont {Parrondo}}, \ and\ \bibinfo
  {author} {\bibfnamefont {R.~A.}\ \bibnamefont {Rica}},\ }\href@noop {}
  {\bibfield  {journal} {\bibinfo  {journal} {Nat. Phys.}\ }\textbf {\bibinfo
  {volume} {12}},\ \bibinfo {pages} {67} (\bibinfo {year} {2016})}\BibitemShut
  {NoStop}%
\bibitem [{\citenamefont {Krishnamurthy}\ \emph {et~al.}(2016)\citenamefont
  {Krishnamurthy}, \citenamefont {Ghosh}, \citenamefont {Chatterji},
  \citenamefont {Ganapathy},\ and\ \citenamefont {Sood}}]{Sood2016}%
  \BibitemOpen
  \bibfield  {author} {\bibinfo {author} {\bibfnamefont {S.}~\bibnamefont
  {Krishnamurthy}}, \bibinfo {author} {\bibfnamefont {S.}~\bibnamefont
  {Ghosh}}, \bibinfo {author} {\bibfnamefont {D.}~\bibnamefont {Chatterji}},
  \bibinfo {author} {\bibfnamefont {R.}~\bibnamefont {Ganapathy}}, \ and\
  \bibinfo {author} {\bibfnamefont {A.~K.}\ \bibnamefont {Sood}},\ }\href@noop
  {} {\bibfield  {journal} {\bibinfo  {journal} {Nat. Phys.}\ }\textbf
  {\bibinfo {volume} {12}},\ \bibinfo {pages} {1134} (\bibinfo {year}
  {2016})}\BibitemShut {NoStop}%
\bibitem [{\citenamefont {Yan}\ \emph {et~al.}(2014)\citenamefont {Yan},
  \citenamefont {Gray},\ and\ \citenamefont {Scherer}}]{Norbert2014}%
  \BibitemOpen
  \bibfield  {author} {\bibinfo {author} {\bibfnamefont {Z.}~\bibnamefont
  {Yan}}, \bibinfo {author} {\bibfnamefont {S.~K.}\ \bibnamefont {Gray}}, \
  and\ \bibinfo {author} {\bibfnamefont {N.~F.}\ \bibnamefont {Scherer}},\
  }\href@noop {} {\bibfield  {journal} {\bibinfo  {journal} {Nat. Comm.}\
  }\textbf {\bibinfo {volume} {5}},\ \bibinfo {pages} {3751} (\bibinfo {year}
  {2014})}\BibitemShut {NoStop}%
\bibitem [{\citenamefont {Howse}\ \emph {et~al.}(2007)\citenamefont {Howse},
  \citenamefont {Jones}, \citenamefont {Ryan}, \citenamefont {Gough},
  \citenamefont {Vafabakhsh},\ and\ \citenamefont
  {Golestanian}}]{Golestanian2007}%
  \BibitemOpen
  \bibfield  {author} {\bibinfo {author} {\bibfnamefont {J.~R.}\ \bibnamefont
  {Howse}}, \bibinfo {author} {\bibfnamefont {R.~A.~L.}\ \bibnamefont {Jones}},
  \bibinfo {author} {\bibfnamefont {A.~J.}\ \bibnamefont {Ryan}}, \bibinfo
  {author} {\bibfnamefont {T.}~\bibnamefont {Gough}}, \bibinfo {author}
  {\bibfnamefont {R.}~\bibnamefont {Vafabakhsh}}, \ and\ \bibinfo {author}
  {\bibfnamefont {R.}~\bibnamefont {Golestanian}},\ }\href {\doibase
  10.1103/PhysRevLett.99.048102} {\bibfield  {journal} {\bibinfo  {journal}
  {Phys. Rev. Lett.}\ }\textbf {\bibinfo {volume} {99}},\ \bibinfo {pages}
  {048102} (\bibinfo {year} {2007})}\BibitemShut {NoStop}%
\bibitem [{\citenamefont {Prigogine}(1978)}]{Prigogine1978}%
  \BibitemOpen
  \bibfield  {author} {\bibinfo {author} {\bibfnamefont {I.}~\bibnamefont
  {Prigogine}},\ }\href@noop {} {\bibfield  {journal} {\bibinfo  {journal}
  {Science}\ }\textbf {\bibinfo {volume} {201}},\ \bibinfo {pages} {777}
  (\bibinfo {year} {1978})}\BibitemShut {NoStop}%
\bibitem [{\citenamefont {Kadanoff}(2001)}]{Kadanoff2001}%
  \BibitemOpen
  \bibfield  {author} {\bibinfo {author} {\bibfnamefont {L.~P.}\ \bibnamefont
  {Kadanoff}},\ }\href@noop {} {\bibfield  {journal} {\bibinfo  {journal}
  {Physics Today}\ }\textbf {\bibinfo {volume} {54}},\ \bibinfo {pages} {34}
  (\bibinfo {year} {2001})}\BibitemShut {NoStop}%
\bibitem [{\citenamefont {Grosberg}\ and\ \citenamefont
  {Joanny}(2015)}]{Grosberg2015}%
  \BibitemOpen
  \bibfield  {author} {\bibinfo {author} {\bibfnamefont {A.~Y.}\ \bibnamefont
  {Grosberg}}\ and\ \bibinfo {author} {\bibfnamefont {J.-F.}\ \bibnamefont
  {Joanny}},\ }\href {\doibase 10.1103/PhysRevE.92.032118} {\bibfield
  {journal} {\bibinfo  {journal} {Phys. Rev. E}\ }\textbf {\bibinfo {volume}
  {92}},\ \bibinfo {pages} {032118} (\bibinfo {year} {2015})}\BibitemShut
  {NoStop}%
\bibitem [{\citenamefont {Weber}\ \emph {et~al.}(2016)\citenamefont {Weber},
  \citenamefont {Weber},\ and\ \citenamefont {Frey}}]{Weber2016}%
  \BibitemOpen
  \bibfield  {author} {\bibinfo {author} {\bibfnamefont {S.~N.}\ \bibnamefont
  {Weber}}, \bibinfo {author} {\bibfnamefont {C.~A.}\ \bibnamefont {Weber}}, \
  and\ \bibinfo {author} {\bibfnamefont {E.}~\bibnamefont {Frey}},\ }\href@noop
  {} {\bibfield  {journal} {\bibinfo  {journal} {Phys. Rev. Lett.}\ }\textbf
  {\bibinfo {volume} {116}},\ \bibinfo {pages} {058301} (\bibinfo {year}
  {2016})}\BibitemShut {NoStop}%
\bibitem [{\citenamefont {Cates}\ and\ \citenamefont
  {Tailleur}(2015)}]{Cates2016}%
  \BibitemOpen
  \bibfield  {author} {\bibinfo {author} {\bibfnamefont {M.~E.}\ \bibnamefont
  {Cates}}\ and\ \bibinfo {author} {\bibfnamefont {J.}~\bibnamefont
  {Tailleur}},\ }\href@noop {} {\bibfield  {journal} {\bibinfo  {journal}
  {Annu. Rev. Condens. Matter Phys.}\ }\textbf {\bibinfo {volume} {6}},\
  \bibinfo {pages} {219} (\bibinfo {year} {2015})}\BibitemShut {NoStop}%
\bibitem [{\citenamefont {Dotsenko}\ \emph {et~al.}(2013)\citenamefont
  {Dotsenko}, \citenamefont {Maciolek}, \citenamefont {Vasilyev},\ and\
  \citenamefont {Oshanin}}]{Oshanin2013}%
  \BibitemOpen
  \bibfield  {author} {\bibinfo {author} {\bibfnamefont {V.}~\bibnamefont
  {Dotsenko}}, \bibinfo {author} {\bibfnamefont {A.}~\bibnamefont {Maciolek}},
  \bibinfo {author} {\bibfnamefont {O.}~\bibnamefont {Vasilyev}}, \ and\
  \bibinfo {author} {\bibfnamefont {G.}~\bibnamefont {Oshanin}},\ }\href
  {\doibase 10.1103/PhysRevE.87.062130} {\bibfield  {journal} {\bibinfo
  {journal} {Phys. Rev. E}\ }\textbf {\bibinfo {volume} {87}},\ \bibinfo
  {pages} {062130} (\bibinfo {year} {2013})}\BibitemShut {NoStop}%
\bibitem [{\citenamefont {Dotsenko}\ \emph {et~al.}(2019)\citenamefont
  {Dotsenko}, \citenamefont {Maciolek}, \citenamefont {Oshanin}, \citenamefont
  {Vasilyev},\ and\ \citenamefont {Dietrich}}]{Oshanin2019}%
  \BibitemOpen
  \bibfield  {author} {\bibinfo {author} {\bibfnamefont {V.~S.}\ \bibnamefont
  {Dotsenko}}, \bibinfo {author} {\bibfnamefont {A.}~\bibnamefont {Maciolek}},
  \bibinfo {author} {\bibfnamefont {G.}~\bibnamefont {Oshanin}}, \bibinfo
  {author} {\bibfnamefont {O.}~\bibnamefont {Vasilyev}}, \ and\ \bibinfo
  {author} {\bibfnamefont {S.}~\bibnamefont {Dietrich}},\ }\href@noop {}
  {\bibfield  {journal} {\bibinfo  {journal} {New J. Phys.}\ }\textbf {\bibinfo
  {volume} {21}},\ \bibinfo {pages} {033036} (\bibinfo {year}
  {2019})}\BibitemShut {NoStop}%
\bibitem [{\citenamefont {Grosberg}\ and\ \citenamefont
  {Joanny}(2018)}]{Grosberg2018}%
  \BibitemOpen
  \bibfield  {author} {\bibinfo {author} {\bibfnamefont {A.~Y.}\ \bibnamefont
  {Grosberg}}\ and\ \bibinfo {author} {\bibfnamefont {J.-F.}\ \bibnamefont
  {Joanny}},\ }\href@noop {} {\bibfield  {journal} {\bibinfo  {journal} {Pol.
  Sci., Ser. C}\ }\textbf {\bibinfo {volume} {60}},\ \bibinfo {pages} {118}
  (\bibinfo {year} {2018})}\BibitemShut {NoStop}%
\bibitem [{\citenamefont {Battle}\ \emph {et~al.}(2016)\citenamefont {Battle},
  \citenamefont {Broedersz}, \citenamefont {Fakhri}, \citenamefont {Geyer},
  \citenamefont {Howard}, \citenamefont {Schmidt},\ and\ \citenamefont
  {MacKintosh}}]{Battle2016}%
  \BibitemOpen
  \bibfield  {author} {\bibinfo {author} {\bibfnamefont {C.}~\bibnamefont
  {Battle}}, \bibinfo {author} {\bibfnamefont {C.~P.}\ \bibnamefont
  {Broedersz}}, \bibinfo {author} {\bibfnamefont {N.}~\bibnamefont {Fakhri}},
  \bibinfo {author} {\bibfnamefont {V.~F.}\ \bibnamefont {Geyer}}, \bibinfo
  {author} {\bibfnamefont {J.}~\bibnamefont {Howard}}, \bibinfo {author}
  {\bibfnamefont {C.~F.}\ \bibnamefont {Schmidt}}, \ and\ \bibinfo {author}
  {\bibfnamefont {F.~C.}\ \bibnamefont {MacKintosh}},\ }\href {\doibase
  10.1126/science.aac8167} {\bibfield  {journal} {\bibinfo  {journal}
  {Science}\ }\textbf {\bibinfo {volume} {352}},\ \bibinfo {pages} {604}
  (\bibinfo {year} {2016})}\BibitemShut {NoStop}%
\bibitem [{\citenamefont {Li}\ \emph {et~al.}(2019)\citenamefont {Li},
  \citenamefont {Horowitz}, \citenamefont {Gingrich},\ and\ \citenamefont
  {Fakhri}}]{ref_Nat_Comm}%
  \BibitemOpen
  \bibfield  {author} {\bibinfo {author} {\bibfnamefont {J.}~\bibnamefont
  {Li}}, \bibinfo {author} {\bibfnamefont {J.~M.}\ \bibnamefont {Horowitz}},
  \bibinfo {author} {\bibfnamefont {T.~R.}\ \bibnamefont {Gingrich}}, \ and\
  \bibinfo {author} {\bibfnamefont {N.}~\bibnamefont {Fakhri}},\ }\href@noop {}
  {\bibfield  {journal} {\bibinfo  {journal} {Nature Communications}\ }\textbf
  {\bibinfo {volume} {10}},\ \bibinfo {pages} {1666} (\bibinfo {year}
  {2019})}\BibitemShut {NoStop}%
\bibitem [{\citenamefont {Wang}\ and\ \citenamefont
  {Grosberg}(2020)}]{different_temp_Grosberg}%
  \BibitemOpen
  \bibfield  {author} {\bibinfo {author} {\bibfnamefont {M.}~\bibnamefont
  {Wang}}\ and\ \bibinfo {author} {\bibfnamefont {A.~Y.}\ \bibnamefont
  {Grosberg}},\ }\href@noop {} {\bibfield  {journal} {\bibinfo  {journal}
  {Phys. Rev. E}\ }\textbf {\bibinfo {volume} {101}},\ \bibinfo {pages}
  {032131} (\bibinfo {year} {2020})}\BibitemShut {NoStop}%
\bibitem [{\citenamefont {Tanaka}\ \emph {et~al.}(2017)\citenamefont {Tanaka},
  \citenamefont {Lee},\ and\ \citenamefont {Brenner}}]{PhysRevFluids_harvard}%
  \BibitemOpen
  \bibfield  {author} {\bibinfo {author} {\bibfnamefont {H.}~\bibnamefont
  {Tanaka}}, \bibinfo {author} {\bibfnamefont {A.~A.}\ \bibnamefont {Lee}}, \
  and\ \bibinfo {author} {\bibfnamefont {M.~P.}\ \bibnamefont {Brenner}},\
  }\href@noop {} {\bibfield  {journal} {\bibinfo  {journal} {Phys. Rev.
  Fluids}\ }\textbf {\bibinfo {volume} {2}},\ \bibinfo {pages} {043103}
  (\bibinfo {year} {2017})}\BibitemShut {NoStop}%
\bibitem [{\citenamefont {Wang}\ \emph {et~al.}(2021)\citenamefont {Wang},
  \citenamefont {Zinga}, \citenamefont {Zidovska},\ and\ \citenamefont
  {Grosberg}}]{Grosberg_2021}%
  \BibitemOpen
  \bibfield  {author} {\bibinfo {author} {\bibfnamefont {M.}~\bibnamefont
  {Wang}}, \bibinfo {author} {\bibfnamefont {K.}~\bibnamefont {Zinga}},
  \bibinfo {author} {\bibfnamefont {A.}~\bibnamefont {Zidovska}}, \ and\
  \bibinfo {author} {\bibfnamefont {A.~Y.}\ \bibnamefont {Grosberg}},\
  }\href@noop {} {\bibfield  {journal} {\bibinfo  {journal} {Soft Matter}\
  }\textbf {\bibinfo {volume} {17}},\ \bibinfo {pages} {9528} (\bibinfo {year}
  {2021})}\BibitemShut {NoStop}%
\bibitem [{\citenamefont {Jardat}\ \emph {et~al.}(2022)\citenamefont {Jardat},
  \citenamefont {Dahirel},\ and\ \citenamefont
  {Illien}}]{soft_particles_different_thermostats}%
  \BibitemOpen
  \bibfield  {author} {\bibinfo {author} {\bibfnamefont {M.}~\bibnamefont
  {Jardat}}, \bibinfo {author} {\bibfnamefont {V.}~\bibnamefont {Dahirel}}, \
  and\ \bibinfo {author} {\bibfnamefont {P.}~\bibnamefont {Illien}},\
  }\href@noop {} {\bibfield  {journal} {\bibinfo  {journal} {Phys. Rev. E}\
  }\textbf {\bibinfo {volume} {106}},\ \bibinfo {pages} {064608} (\bibinfo
  {year} {2022})}\BibitemShut {NoStop}%
\bibitem [{\citenamefont {Gardiner}(2004)}]{Gardiner}%
  \BibitemOpen
  \bibfield  {author} {\bibinfo {author} {\bibfnamefont {C.}~\bibnamefont
  {Gardiner}},\ }\href@noop {} {\emph {\bibinfo {title} {Handbook of stochastic
  methods for physics, chemistry and the natural sciences}}}\ (\bibinfo
  {publisher} {Springer, Berlin},\ \bibinfo {year} {2004})\BibitemShut
  {NoStop}%
\bibitem [{\citenamefont {Chandler}\ \emph {et~al.}(1983)\citenamefont
  {Chandler}, \citenamefont {Weeks},\ and\ \citenamefont
  {Andersen}}]{Chandler_WCA}%
  \BibitemOpen
  \bibfield  {author} {\bibinfo {author} {\bibfnamefont {D.}~\bibnamefont
  {Chandler}}, \bibinfo {author} {\bibfnamefont {J.~D.}\ \bibnamefont {Weeks}},
  \ and\ \bibinfo {author} {\bibfnamefont {H.~C.}\ \bibnamefont {Andersen}},\
  }\href@noop {} {\bibfield  {journal} {\bibinfo  {journal} {Science}\ }\textbf
  {\bibinfo {volume} {220}},\ \bibinfo {pages} {787} (\bibinfo {year}
  {1983})}\BibitemShut {NoStop}%
\bibitem [{\citenamefont {Hidalgo-Soria}\ and\ \citenamefont
  {Barkai}(2021)}]{Mario2021}%
  \BibitemOpen
  \bibfield  {author} {\bibinfo {author} {\bibfnamefont {M.}~\bibnamefont
  {Hidalgo-Soria}}\ and\ \bibinfo {author} {\bibfnamefont {S.}~\bibnamefont
  {Barkai}, \bibfnamefont {E.amd~Burov}},\ }\href@noop {} {\bibfield  {journal}
  {\bibinfo  {journal} {Entropy}\ }\textbf {\bibinfo {volume} {23}},\ \bibinfo
  {pages} {231} (\bibinfo {year} {2021})}\BibitemShut {NoStop}%
\bibitem [{\citenamefont {Mandal}\ \emph {et~al.}(2020)\citenamefont {Mandal},
  \citenamefont {Kurzthaler}, \citenamefont {Franosch},\ and\ \citenamefont
  {L\"owen}}]{Lowen2020}%
  \BibitemOpen
  \bibfield  {author} {\bibinfo {author} {\bibfnamefont {S.}~\bibnamefont
  {Mandal}}, \bibinfo {author} {\bibfnamefont {C.}~\bibnamefont {Kurzthaler}},
  \bibinfo {author} {\bibfnamefont {T.}~\bibnamefont {Franosch}}, \ and\
  \bibinfo {author} {\bibfnamefont {H.}~\bibnamefont {L\"owen}},\ }\href
  {\doibase 10.1103/PhysRevLett.125.138002} {\bibfield  {journal} {\bibinfo
  {journal} {Phys. Rev. Lett.}\ }\textbf {\bibinfo {volume} {125}},\ \bibinfo
  {pages} {138002} (\bibinfo {year} {2020})}\BibitemShut {NoStop}%
\bibitem [{\citenamefont {Agudo-Canalejo}\ \emph {et~al.}(2018)\citenamefont
  {Agudo-Canalejo}, \citenamefont {Illien},\ and\ \citenamefont
  {Golestanian}}]{Golestenian2018}%
  \BibitemOpen
  \bibfield  {author} {\bibinfo {author} {\bibfnamefont {J.}~\bibnamefont
  {Agudo-Canalejo}}, \bibinfo {author} {\bibfnamefont {P.}~\bibnamefont
  {Illien}}, \ and\ \bibinfo {author} {\bibfnamefont {R.}~\bibnamefont
  {Golestanian}},\ }\href@noop {} {\bibfield  {journal} {\bibinfo  {journal}
  {Nano Lett.}\ }\textbf {\bibinfo {volume} {18}},\ \bibinfo {pages} {2711}
  (\bibinfo {year} {2018})}\BibitemShut {NoStop}%
\bibitem [{\citenamefont {Wang}\ \emph {et~al.}(2020)\citenamefont {Wang},
  \citenamefont {Park}, \citenamefont {Dong}, \citenamefont {Kim},
  \citenamefont {Cho}, \citenamefont {Tlusty},\ and\ \citenamefont
  {Granick}}]{Granick2020A}%
  \BibitemOpen
  \bibfield  {author} {\bibinfo {author} {\bibfnamefont {H.}~\bibnamefont
  {Wang}}, \bibinfo {author} {\bibfnamefont {M.}~\bibnamefont {Park}}, \bibinfo
  {author} {\bibfnamefont {R.}~\bibnamefont {Dong}}, \bibinfo {author}
  {\bibfnamefont {J.}~\bibnamefont {Kim}}, \bibinfo {author} {\bibfnamefont
  {Y.-K.}\ \bibnamefont {Cho}}, \bibinfo {author} {\bibfnamefont
  {T.}~\bibnamefont {Tlusty}}, \ and\ \bibinfo {author} {\bibfnamefont
  {S.}~\bibnamefont {Granick}},\ }\href {\doibase 10.1126/science.aba8425}
  {\bibfield  {journal} {\bibinfo  {journal} {Science}\ }\textbf {\bibinfo
  {volume} {369}},\ \bibinfo {pages} {537} (\bibinfo {year}
  {2020})}\BibitemShut {NoStop}%
\bibitem [{\citenamefont {Jee}\ \emph {et~al.}(2020)\citenamefont {Jee},
  \citenamefont {Tlusty},\ and\ \citenamefont {Granick}}]{Granick2020B}%
  \BibitemOpen
  \bibfield  {author} {\bibinfo {author} {\bibfnamefont {A.-Y.}\ \bibnamefont
  {Jee}}, \bibinfo {author} {\bibfnamefont {T.}~\bibnamefont {Tlusty}}, \ and\
  \bibinfo {author} {\bibfnamefont {S.}~\bibnamefont {Granick}},\ }\href
  {\doibase 10.1073/pnas.2019810117} {\bibfield  {journal} {\bibinfo  {journal}
  {Proc. Nat. Acad. Sci.}\ }\textbf {\bibinfo {volume} {117}},\ \bibinfo
  {pages} {29435} (\bibinfo {year} {2020})}\BibitemShut {NoStop}%
\bibitem [{\citenamefont {Feng}\ and\ \citenamefont
  {Gilson}(2020)}]{Gilson2020}%
  \BibitemOpen
  \bibfield  {author} {\bibinfo {author} {\bibfnamefont {M.}~\bibnamefont
  {Feng}}\ and\ \bibinfo {author} {\bibfnamefont {M.~K.}\ \bibnamefont
  {Gilson}},\ }\href@noop {} {\bibfield  {journal} {\bibinfo  {journal} {Annu.
  Rev. Biophys.}\ }\textbf {\bibinfo {volume} {49}},\ \bibinfo {pages} {87}
  (\bibinfo {year} {2020})}\BibitemShut {NoStop}%
\bibitem [{\citenamefont {Scala}\ \emph {et~al.}(2000)\citenamefont {Scala},
  \citenamefont {Starr}, \citenamefont {La~Nave}, \citenamefont {Sciortino},\
  and\ \citenamefont {Stanley}}]{Stanely200}%
  \BibitemOpen
  \bibfield  {author} {\bibinfo {author} {\bibfnamefont {A.}~\bibnamefont
  {Scala}}, \bibinfo {author} {\bibfnamefont {F.}~\bibnamefont {Starr}},
  \bibinfo {author} {\bibfnamefont {E.}~\bibnamefont {La~Nave}}, \bibinfo
  {author} {\bibfnamefont {F.}~\bibnamefont {Sciortino}}, \ and\ \bibinfo
  {author} {\bibfnamefont {H.~E.}\ \bibnamefont {Stanley}},\ }\href@noop {}
  {\bibfield  {journal} {\bibinfo  {journal} {Nature}\ }\textbf {\bibinfo
  {volume} {406}},\ \bibinfo {pages} {166} (\bibinfo {year}
  {2000})}\BibitemShut {NoStop}%
\bibitem [{\citenamefont {Debenedetti}\ and\ \citenamefont
  {Stillinger}(2001)}]{Stillinger2001}%
  \BibitemOpen
  \bibfield  {author} {\bibinfo {author} {\bibfnamefont {P.~G.}\ \bibnamefont
  {Debenedetti}}\ and\ \bibinfo {author} {\bibfnamefont {F.~H.}\ \bibnamefont
  {Stillinger}},\ }\href@noop {} {\bibfield  {journal} {\bibinfo  {journal}
  {Nature}\ }\textbf {\bibinfo {volume} {410}},\ \bibinfo {pages} {259}
  (\bibinfo {year} {2001})}\BibitemShut {NoStop}%
\bibitem [{\citenamefont {Charbonneau}\ \emph {et~al.}(2014)\citenamefont
  {Charbonneau}, \citenamefont {Jin}, \citenamefont {Parisi},\ and\
  \citenamefont {Zamponi}}]{Zamponi2014}%
  \BibitemOpen
  \bibfield  {author} {\bibinfo {author} {\bibfnamefont {P.}~\bibnamefont
  {Charbonneau}}, \bibinfo {author} {\bibfnamefont {Y.}~\bibnamefont {Jin}},
  \bibinfo {author} {\bibfnamefont {G.}~\bibnamefont {Parisi}}, \ and\ \bibinfo
  {author} {\bibfnamefont {F.}~\bibnamefont {Zamponi}},\ }\href@noop {}
  {\bibfield  {journal} {\bibinfo  {journal} {Proc. Nat. Acad. Sci.}\ }\textbf
  {\bibinfo {volume} {111}},\ \bibinfo {pages} {15025} (\bibinfo {year}
  {2014})}\BibitemShut {NoStop}%
\bibitem [{\citenamefont {Wei}\ \emph {et~al.}(2018)\citenamefont {Wei},
  \citenamefont {Evenson}, \citenamefont {Stolpe}, \citenamefont {Lucas},\ and\
  \citenamefont {Angell}}]{Angell2018}%
  \BibitemOpen
  \bibfield  {author} {\bibinfo {author} {\bibfnamefont {S.}~\bibnamefont
  {Wei}}, \bibinfo {author} {\bibfnamefont {Z.}~\bibnamefont {Evenson}},
  \bibinfo {author} {\bibfnamefont {M.}~\bibnamefont {Stolpe}}, \bibinfo
  {author} {\bibfnamefont {P.}~\bibnamefont {Lucas}}, \ and\ \bibinfo {author}
  {\bibfnamefont {C.~A.}\ \bibnamefont {Angell}},\ }\href@noop {} {\bibfield
  {journal} {\bibinfo  {journal} {Sci. Adv.}\ }\textbf {\bibinfo {volume}
  {4}},\ \bibinfo {pages} {1} (\bibinfo {year} {2018})}\BibitemShut {NoStop}%
\bibitem [{\citenamefont {Doliwa}\ and\ \citenamefont
  {Heuer}(1998)}]{uncoraleted_Heuer}%
  \BibitemOpen
  \bibfield  {author} {\bibinfo {author} {\bibfnamefont {B.}~\bibnamefont
  {Doliwa}}\ and\ \bibinfo {author} {\bibfnamefont {A.}~\bibnamefont {Heuer}},\
  }\href@noop {} {\bibfield  {journal} {\bibinfo  {journal} {Phys. Rev. Lett.}\
  }\textbf {\bibinfo {volume} {80}},\ \bibinfo {pages} {4915} (\bibinfo {year}
  {1998})}\BibitemShut {NoStop}%
\bibitem [{\citenamefont {Cugliandolo}(2011)}]{Cugliandolo_effective_temp}%
  \BibitemOpen
  \bibfield  {author} {\bibinfo {author} {\bibfnamefont {L.~F.}\ \bibnamefont
  {Cugliandolo}},\ }\href@noop {} {\bibfield  {journal} {\bibinfo  {journal}
  {Journal of Physics A: Mathematical and Theoretical}\ }\textbf {\bibinfo
  {volume} {44}},\ \bibinfo {pages} {483001} (\bibinfo {year}
  {2011})}\BibitemShut {NoStop}%
\bibitem [{\citenamefont {L{\'{e}}onard}\ and\ \citenamefont
  {Berthier}(2005)}]{L_onard_2005}%
  \BibitemOpen
  \bibfield  {author} {\bibinfo {author} {\bibfnamefont {S.}~\bibnamefont
  {L{\'{e}}onard}}\ and\ \bibinfo {author} {\bibfnamefont {L.}~\bibnamefont
  {Berthier}},\ }\href@noop {} {\bibfield  {journal} {\bibinfo  {journal}
  {Journal of Physics: Condensed Matter}\ }\textbf {\bibinfo {volume} {17}},\
  \bibinfo {pages} {S3571} (\bibinfo {year} {2005})}\BibitemShut {NoStop}%
\bibitem [{\citenamefont {Heuer}\ and\ \citenamefont
  {Okun}(1997)}]{Hetero_homo_dynamics_Heuer}%
  \BibitemOpen
  \bibfield  {author} {\bibinfo {author} {\bibfnamefont {A.}~\bibnamefont
  {Heuer}}\ and\ \bibinfo {author} {\bibfnamefont {K.}~\bibnamefont {Okun}},\
  }\href@noop {} {\bibfield  {journal} {\bibinfo  {journal} {The Journal of
  Chemical Physics}\ }\textbf {\bibinfo {volume} {106}},\ \bibinfo {pages}
  {6176} (\bibinfo {year} {1997})}\BibitemShut {NoStop}%
\bibitem [{\citenamefont {Burov}\ \emph {et~al.}(2013)\citenamefont {Burov},
  \citenamefont {Tabei}, \citenamefont {Huynh}, \citenamefont {Murrell},
  \citenamefont {Philipson}, \citenamefont {Rice}, \citenamefont {Gardel},
  \citenamefont {Scherer},\ and\ \citenamefont {Dinner}}]{Burov2013}%
  \BibitemOpen
  \bibfield  {author} {\bibinfo {author} {\bibfnamefont {S.}~\bibnamefont
  {Burov}}, \bibinfo {author} {\bibfnamefont {S.~M.~A.}\ \bibnamefont {Tabei}},
  \bibinfo {author} {\bibfnamefont {T.}~\bibnamefont {Huynh}}, \bibinfo
  {author} {\bibfnamefont {M.~P.}\ \bibnamefont {Murrell}}, \bibinfo {author}
  {\bibfnamefont {L.~H.}\ \bibnamefont {Philipson}}, \bibinfo {author}
  {\bibfnamefont {S.~A.}\ \bibnamefont {Rice}}, \bibinfo {author}
  {\bibfnamefont {M.~L.}\ \bibnamefont {Gardel}}, \bibinfo {author}
  {\bibfnamefont {N.~F.}\ \bibnamefont {Scherer}}, \ and\ \bibinfo {author}
  {\bibfnamefont {A.~R.}\ \bibnamefont {Dinner}},\ }\href@noop {} {\bibfield
  {journal} {\bibinfo  {journal} {Proc. Nat. Acad. Sci.}\ }\textbf {\bibinfo
  {volume} {110}},\ \bibinfo {pages} {19689} (\bibinfo {year}
  {2013})}\BibitemShut {NoStop}%
\bibitem [{\citenamefont {Bos}\ \emph {et~al.}(2015)\citenamefont {Bos},
  \citenamefont {Kadoch},\ and\ \citenamefont {Schneider}}]{Schneider2015}%
  \BibitemOpen
  \bibfield  {author} {\bibinfo {author} {\bibfnamefont {W.~J.~T.}\
  \bibnamefont {Bos}}, \bibinfo {author} {\bibfnamefont {B.}~\bibnamefont
  {Kadoch}}, \ and\ \bibinfo {author} {\bibfnamefont {K.}~\bibnamefont
  {Schneider}},\ }\href@noop {} {\bibfield  {journal} {\bibinfo  {journal}
  {Phys. Rev. Lett.}\ }\textbf {\bibinfo {volume} {114}},\ \bibinfo {pages}
  {214502} (\bibinfo {year} {2015})}\BibitemShut {NoStop}%
\bibitem [{\citenamefont {Ariel}\ \emph {et~al.}(2015)\citenamefont {Ariel},
  \citenamefont {Rabani}, \citenamefont {Benisty}, \citenamefont {Partridge},
  \citenamefont {Harshey},\ and\ \citenamefont {Be’er}}]{beer2015}%
  \BibitemOpen
  \bibfield  {author} {\bibinfo {author} {\bibfnamefont {G.}~\bibnamefont
  {Ariel}}, \bibinfo {author} {\bibfnamefont {A.}~\bibnamefont {Rabani}},
  \bibinfo {author} {\bibfnamefont {S.}~\bibnamefont {Benisty}}, \bibinfo
  {author} {\bibfnamefont {J.~D.}\ \bibnamefont {Partridge}}, \bibinfo {author}
  {\bibfnamefont {R.~M.}\ \bibnamefont {Harshey}}, \ and\ \bibinfo {author}
  {\bibfnamefont {A.}~\bibnamefont {Be’er}},\ }\href@noop {} {\bibfield
  {journal} {\bibinfo  {journal} {Nat. Comm.}\ }\textbf {\bibinfo {volume}
  {6}},\ \bibinfo {pages} {8396} (\bibinfo {year} {2015})}\BibitemShut
  {NoStop}%
\bibitem [{\citenamefont {Kadoch}\ \emph {et~al.}(2017)\citenamefont {Kadoch},
  \citenamefont {Bos},\ and\ \citenamefont {Schneider}}]{Schneider2017}%
  \BibitemOpen
  \bibfield  {author} {\bibinfo {author} {\bibfnamefont {B.}~\bibnamefont
  {Kadoch}}, \bibinfo {author} {\bibfnamefont {W.~J.~T.}\ \bibnamefont {Bos}},
  \ and\ \bibinfo {author} {\bibfnamefont {K.}~\bibnamefont {Schneider}},\
  }\href@noop {} {\bibfield  {journal} {\bibinfo  {journal} {Phys. Rev.
  Fluids}\ }\textbf {\bibinfo {volume} {2}},\ \bibinfo {pages} {064604}
  (\bibinfo {year} {2017})}\BibitemShut {NoStop}%
\bibitem [{\citenamefont {Fang}\ \emph {et~al.}(2022)\citenamefont {Fang},
  \citenamefont {Li}, \citenamefont {Guo}, \citenamefont {Liu},\ and\
  \citenamefont {Huang}}]{Huang2022}%
  \BibitemOpen
  \bibfield  {author} {\bibinfo {author} {\bibfnamefont {L.}~\bibnamefont
  {Fang}}, \bibinfo {author} {\bibfnamefont {L.~L.}\ \bibnamefont {Li}},
  \bibinfo {author} {\bibfnamefont {J.~S.}\ \bibnamefont {Guo}}, \bibinfo
  {author} {\bibfnamefont {Y.~W.}\ \bibnamefont {Liu}}, \ and\ \bibinfo
  {author} {\bibfnamefont {X.~R.}\ \bibnamefont {Huang}},\ }\href@noop {}
  {\bibfield  {journal} {\bibinfo  {journal} {Phys. Lett. A}\ }\textbf
  {\bibinfo {volume} {427}} (\bibinfo {year} {2022})}\BibitemShut {NoStop}%
\end{thebibliography}%


\begin{thebibliography}{2}%
\makeatletter
\providecommand \@ifxundefined [1]{%
 \@ifx{#1\undefined}
}%
\providecommand \@ifnum [1]{%
 \ifnum #1\expandafter \@firstoftwo
 \else \expandafter \@secondoftwo
 \fi
}%
\providecommand \@ifx [1]{%
 \ifx #1\expandafter \@firstoftwo
 \else \expandafter \@secondoftwo
 \fi
}%
\providecommand \natexlab [1]{#1}%
\providecommand \enquote  [1]{``#1''}%
\providecommand \bibnamefont  [1]{#1}%
\providecommand \bibfnamefont [1]{#1}%
\providecommand \citenamefont [1]{#1}%
\providecommand \href@noop [0]{\@secondoftwo}%
\providecommand \href [0]{\begingroup \@sanitize@url \@href}%
\providecommand \@href[1]{\@@startlink{#1}\@@href}%
\providecommand \@@href[1]{\endgroup#1\@@endlink}%
\providecommand \@sanitize@url [0]{\catcode `\\12\catcode `\$12\catcode
  `\&12\catcode `\#12\catcode `\^12\catcode `\_12\catcode `\%12\relax}%
\providecommand \@@startlink[1]{}%
\providecommand \@@endlink[0]{}%
\providecommand \url  [0]{\begingroup\@sanitize@url \@url }%
\providecommand \@url [1]{\endgroup\@href {#1}{\urlprefix }}%
\providecommand \urlprefix  [0]{URL }%
\providecommand \Eprint [0]{\href }%
\providecommand \doibase [0]{https://doi.org/}%
\providecommand \selectlanguage [0]{\@gobble}%
\providecommand \bibinfo  [0]{\@secondoftwo}%
\providecommand \bibfield  [0]{\@secondoftwo}%
\providecommand \translation [1]{[#1]}%
\providecommand \BibitemOpen [0]{}%
\providecommand \bibitemStop [0]{}%
\providecommand \bibitemNoStop [0]{.\EOS\space}%
\providecommand \EOS [0]{\spacefactor3000\relax}%
\providecommand \BibitemShut  [1]{\csname bibitem#1\endcsname}%
\let\auto@bib@innerbib\@empty
\bibitem [{\citenamefont {Bechinger}\ \emph {et~al.}(2016)\citenamefont
  {Bechinger}, \citenamefont {Di~Leonardo}, \citenamefont {L\"owen},
  \citenamefont {Reichhardt}, \citenamefont {Volpe},\ and\ \citenamefont
  {Volpe}}]{RevModPhys2016}%
  \BibitemOpen
  \bibfield  {author} {\bibinfo {author} {\bibfnamefont {C.}~\bibnamefont
  {Bechinger}}, \bibinfo {author} {\bibfnamefont {R.}~\bibnamefont
  {Di~Leonardo}}, \bibinfo {author} {\bibfnamefont {H.}~\bibnamefont
  {L\"owen}}, \bibinfo {author} {\bibfnamefont {C.}~\bibnamefont {Reichhardt}},
  \bibinfo {author} {\bibfnamefont {G.}~\bibnamefont {Volpe}},\ and\ \bibinfo
  {author} {\bibfnamefont {G.}~\bibnamefont {Volpe}},\ }\bibfield  {title}
  {\bibinfo {title} {Active particles in complex and crowded environments},\
  }\href {https://doi.org/10.1103/RevModPhys.88.045006} {\bibfield  {journal}
  {\bibinfo  {journal} {Rev. Mod. Phys.}\ }\textbf {\bibinfo {volume} {88}},\
  \bibinfo {pages} {045006} (\bibinfo {year} {2016})}\BibitemShut {NoStop}%
\bibitem [{\citenamefont {Hidalgo-Soria}\ and\ \citenamefont
  {Barkai}(2021)}]{Mario2021}%
  \BibitemOpen
  \bibfield  {author} {\bibinfo {author} {\bibfnamefont {M.}~\bibnamefont
  {Hidalgo-Soria}}\ and\ \bibinfo {author} {\bibfnamefont {S.}~\bibnamefont
  {Barkai}, \bibfnamefont {E.amd~Burov}},\ }\bibfield  {title} {\bibinfo
  {title} {Cusp of non-gaussian density of particles for a diffusing
  diffusivity model},\ }\href@noop {} {\bibfield  {journal} {\bibinfo
  {journal} {Entropy}\ }\textbf {\bibinfo {volume} {23}},\ \bibinfo {pages}
  {231} (\bibinfo {year} {2021})}\BibitemShut {NoStop}%
\end{thebibliography}%

\end{document}



\title{Supplemental Material for: Emergence of Directed Motion in a Crowded Suspension of Overdamped Particles}

\author{Deborah Schwarcz$^1$}
\email{deborah.schwarcz@gmail.com}
\author{Stanislav Burov$^2$}
\email{stasbur@gmail.com}
\affiliation{$^1$Department of Mathematics and $^2$Physics Department, Bar-Ilan 
University, Ramat Gan 5290002,
Israel}

\date{\today}
\begin{abstract}
 Supplemental material includes (I)  the Definition of active Brownian particle. (II)  Simulation results for several ratios of $D_b/D_a$. (III) Details of Figure 2 {\bf(c)} of the main text. (IV) Possible mechanisms of diffusion coefficient enhancement and (V) References to simulation code.
 \end{abstract}
\maketitle


\section{Active Brownian Particle}

In two-dimensions, the motion of active Brownian particle (ABP) is defined as~\cite{RevModPhys2016} 
\begin{equation}
    \begin{array}{c}
    \frac{dx}{dt} = v\, \text{cos}(\phi)+\sqrt{2D_T}\zeta_x,
    \qquad
    \frac{dy}{dt} = v\, \text{sin}(\phi)+\sqrt{2D_T}\zeta_y
    \\
    \frac{d\phi}{dt} = \sqrt{2D_R}\zeta_\phi,
    \end{array}
    \label{eq:abpdef}
\end{equation}
where $D_T$ describes translational diffusion coefficient and $D_R$ is associated with the rotational motion. $v$ is the speed of the active particle and $\zeta_x$,$\zeta_y$, and $\zeta_\phi$ are uncorrelated Gaussian noises.
The mean squared displacement (MSD) of the ABP is provided by
\begin{equation}
\text{MSD}(\Delta) = \left(2D_T+\frac{2v^2}{D_R}\right)\Delta+\frac{2v^2}{D_R^2}\left(e^{-D_R \Delta}-1\right).
    \label{eq:msdabp}
\end{equation}
Notice that $\phi$ is measured in radians therefore the units of $D_R$ are $1/\text{time}$. For short time scales,  the motion is diffusive (linear MSD)
\begin{equation}
\text{MSD}(\Delta)=4D_T \Delta \qquad \Delta<<D_R^{-1},
    \label{eq:msdshort}
\end{equation}
supperdiffusive on intermediate timescales
\begin{equation}
\text{MSD}(\Delta)\sim \Delta^2 \qquad \Delta\sim D_R^{-1},
    \label{eq:msdinter}
\end{equation}
and again diffusive for long time-scales
\begin{equation}
\text{MSD}(\Delta)\sim\left(4D_T+\frac{2v^2}{D_R}\right) \Delta \qquad \Delta >> D_R^{-1}.
    \label{eq:msdlong}
\end{equation}
By comparing Eq.~\eqref{eq:msdabp} to Eq. (2) of the Main Text, we notice that the parameters $A$ and $B$ are related to the speed and $D_R$ of ABP
\begin{equation}
A=\frac{2v^2}{D_R}\qquad B=D_R^{-1}
    \label{eq:maintxtcomp}
\end{equation}

\section{Results for several ratios $D_b/D_a$}
\begin{figure}[t]
\begin{center}$
\begin{array}{lll}
\includegraphics[width=60mm]{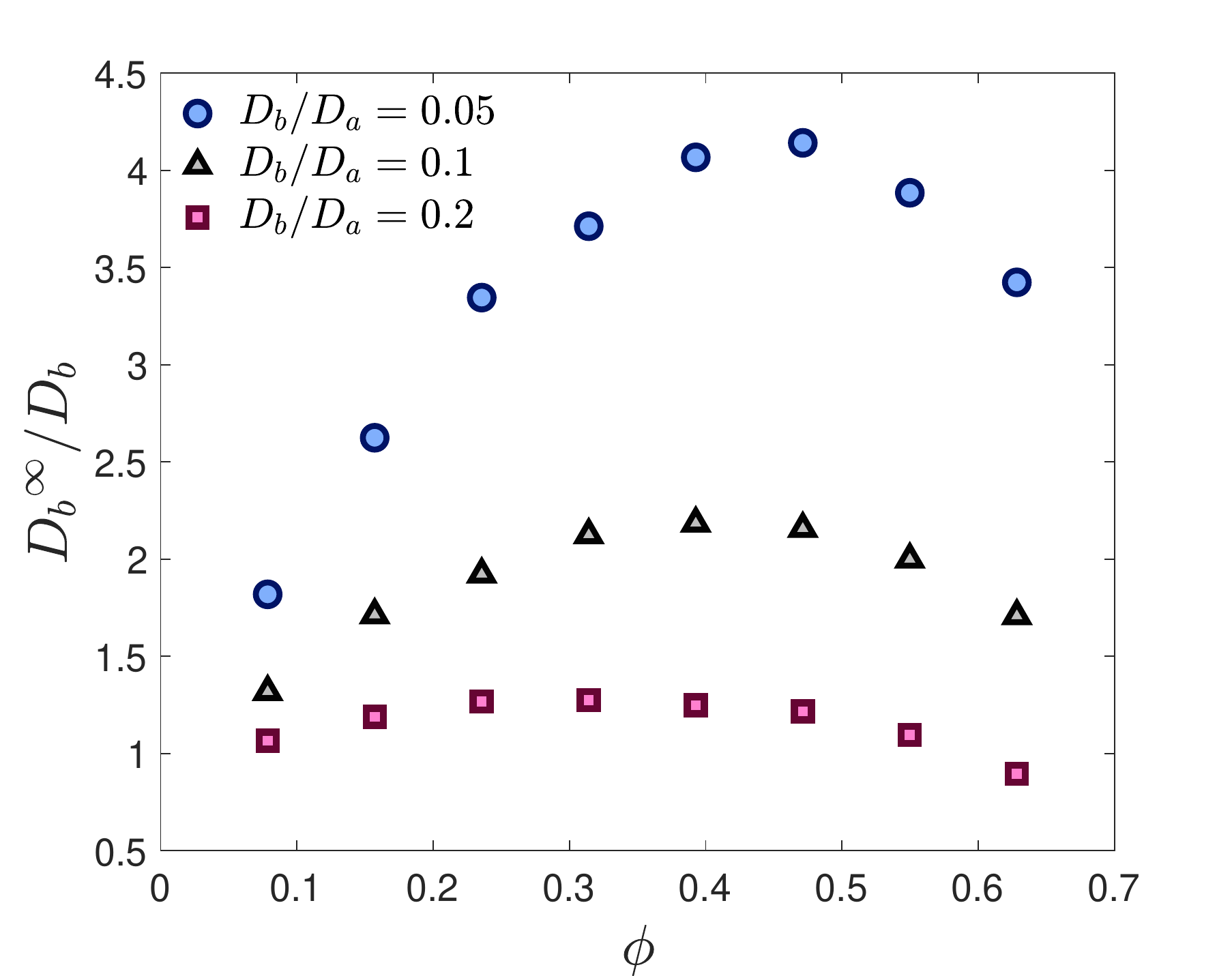} &
	\includegraphics[width=60mm]{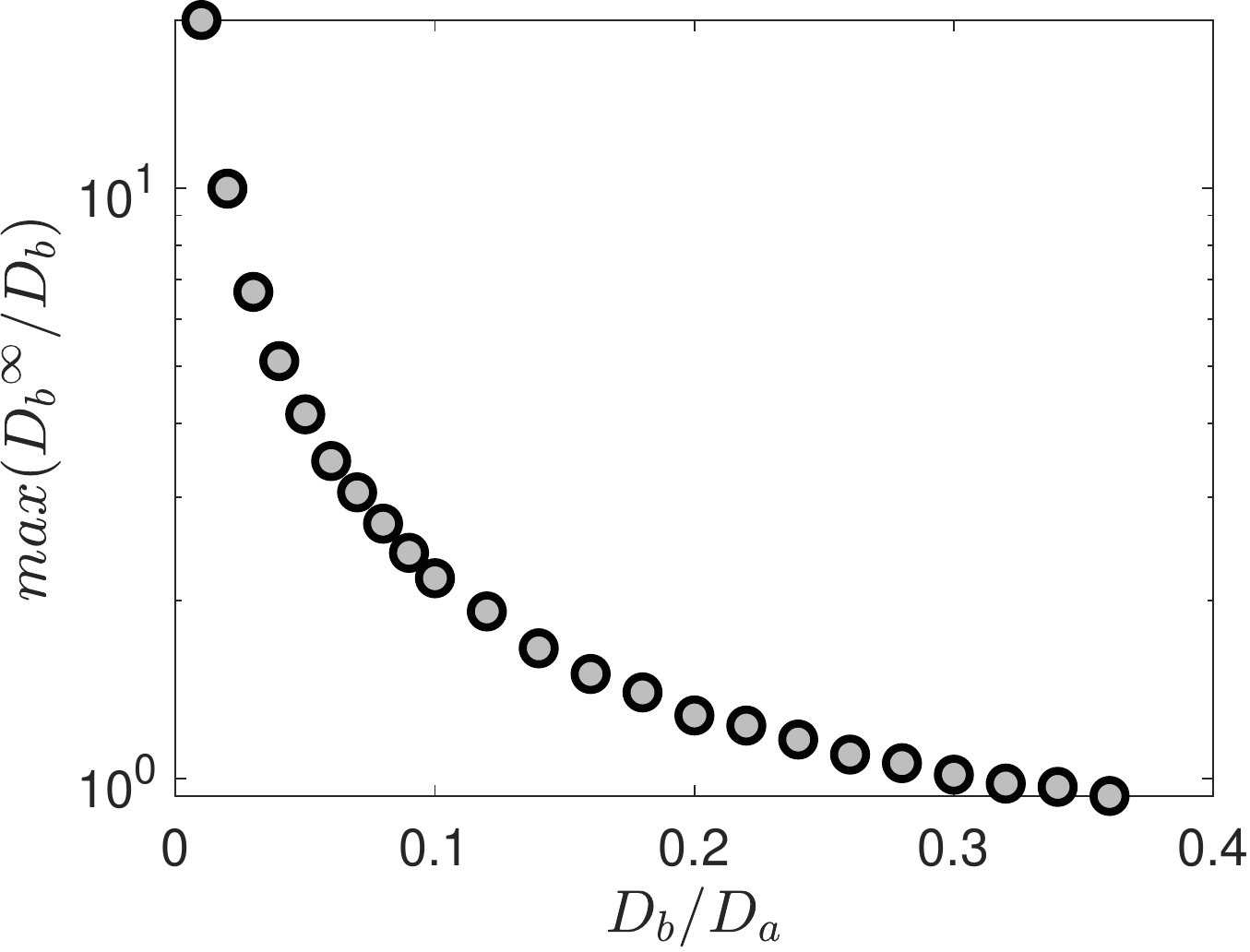} &
\includegraphics[width=60mm]{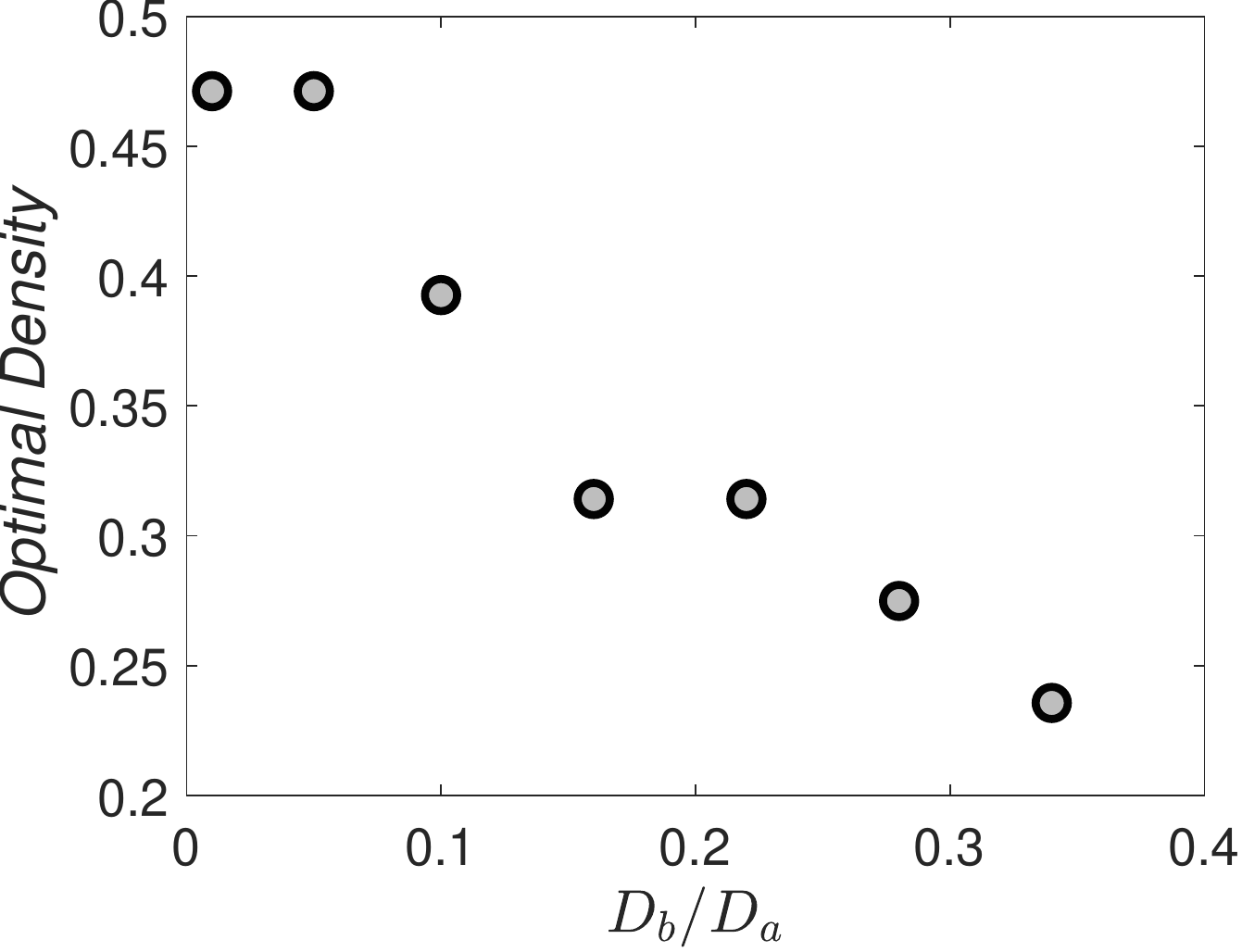}  

\end{array}$
\end{center}
\caption{\textbf {a} $D_b^{\infty}/D_b$ as a function of the density of the solution for $3$ different ratios of diffusivities. \textbf{b} Maximum enhancement of the long-time diffusion coefficient as a function of the ratio $D_b/D_a$. Each data point corresponds to the maximum $D_b^\infty/D_b$ irrespective of the density solution. \textbf {c} Optimal density, i.e., the density for which maximum of $D_b^\infty/D_b$ is obtained in panel (a), as a function of the ratio $D_b/D_a$.
For all panels, $D_a=0.1$, $6\times 10^5$ simulation steps were performed.  Averaging  over time and $40$ ensembles was performed.
}
\label{fig:different_D_b} 
\end{figure}

In the main text, the non-monotonic behavior of the long-time diffusion coefficient $D_b^\infty$ as a function of the density $\phi$ is presented. Fig. 1 {\bf(b)} was obtained for $D_b/D_a=0.1$. In Fig.~\ref{fig:different_D_b} {\bf(a)} we present the behavior of $D_b^\infty$ for two additional ratios $D_b/D_a=0.05$ and $D_b/D_a=0.2$. 
The non-monotonic features repeat themselves. The long-time diffusion coefficient grows as a function of $\phi$ when the density is sufficiently small, and decreases when the density is sufficiently large. 
A new finding is associated with the results presented in Fig.~\ref{fig:different_D_b} {\bf(a)} : the maximal long-time diffusion coefficient $D_b^\infty$ that can be obtained for a given ratio $D_b/D_a$ is decreasing as a function of $D_b/D_a$. 
We present this in Fig.~\ref{fig:different_D_b}. 
Notice that for $D_b/D_a>0.3$ no increase is detected since all the directed motion of the tagged particle starts when $D_b/D_a<0.3$. 
As $D_b\to 0$ the effect becomes more prominent. This is quite reasonable due to the fact that the directed motion is due to the particles with diffusion coefficient $D_a$ that are significantly more mobile as compared to the tagged particle with diffusion coefficient $D_b.$

The maximal value of $D_b^\infty$ is different for every value of the ratio $D_b/D_a$, as we have already seen in Figures \ref{fig:different_D_b} {\bf(a)} and {\bf(b)}.
Another parameter that affects this maximal value of $D_b^\infty$ is the density of the suspension. In Fig.~\ref{fig:different_D_b} {\bf(c)} the optimal density, i.e. the density for which the maximal $D_b^\infty$ is obtained, is plotted as a function of $D_b/D_a$. 
The optimal density increases as $D_b/D_a$ decreases, meaning that the effect of directional motion and the time span of temporally correlated pairs persist for increasingly crowded environments when the ratio $D_b/D_a\to 0$.



\section{Details of Figure 2 {\bf(c)} of the main text}

\begin{figure}[t]
\begin{center}$
\begin{array}{lll}
\includegraphics[width=60mm]{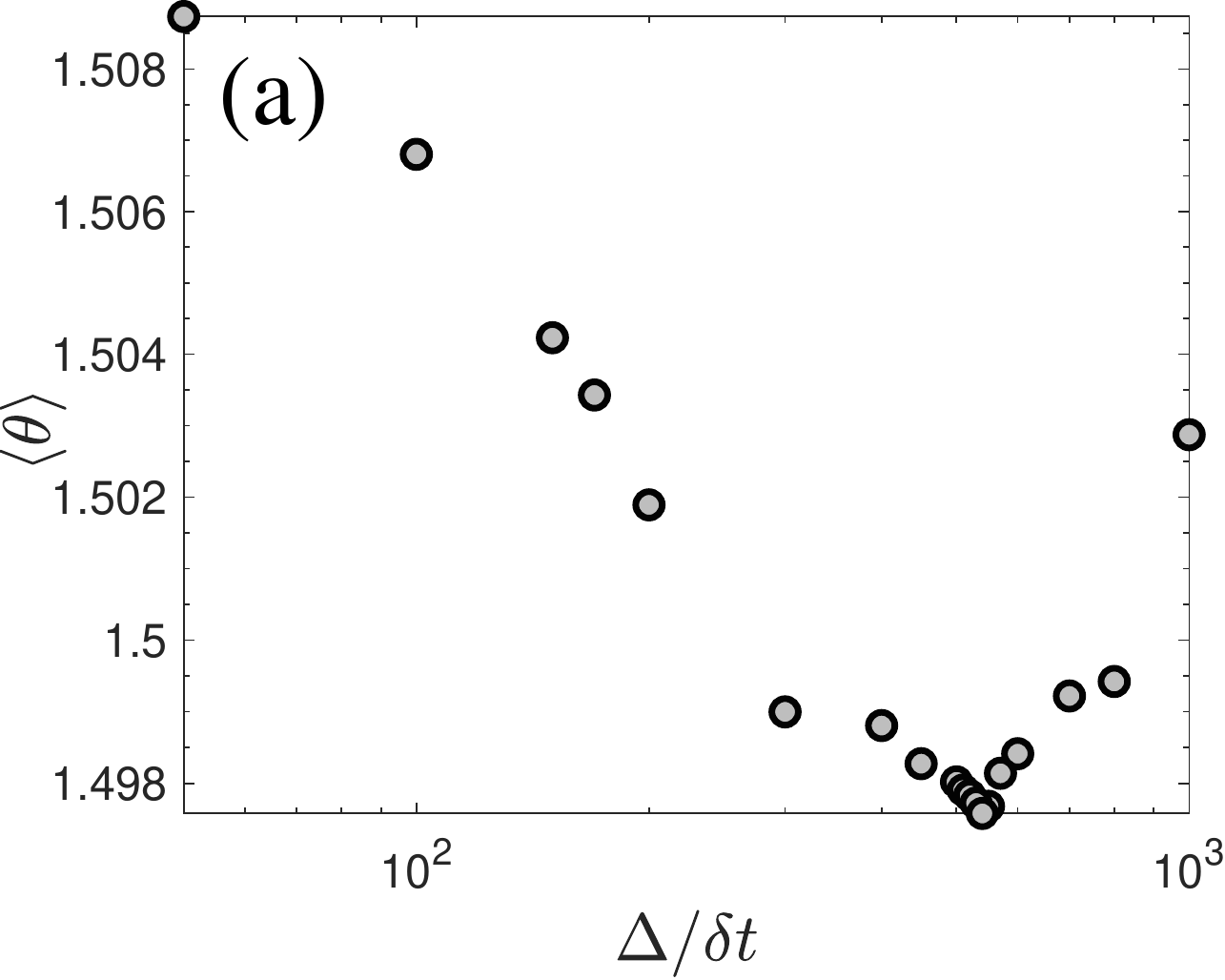} &
\includegraphics[width=63mm]{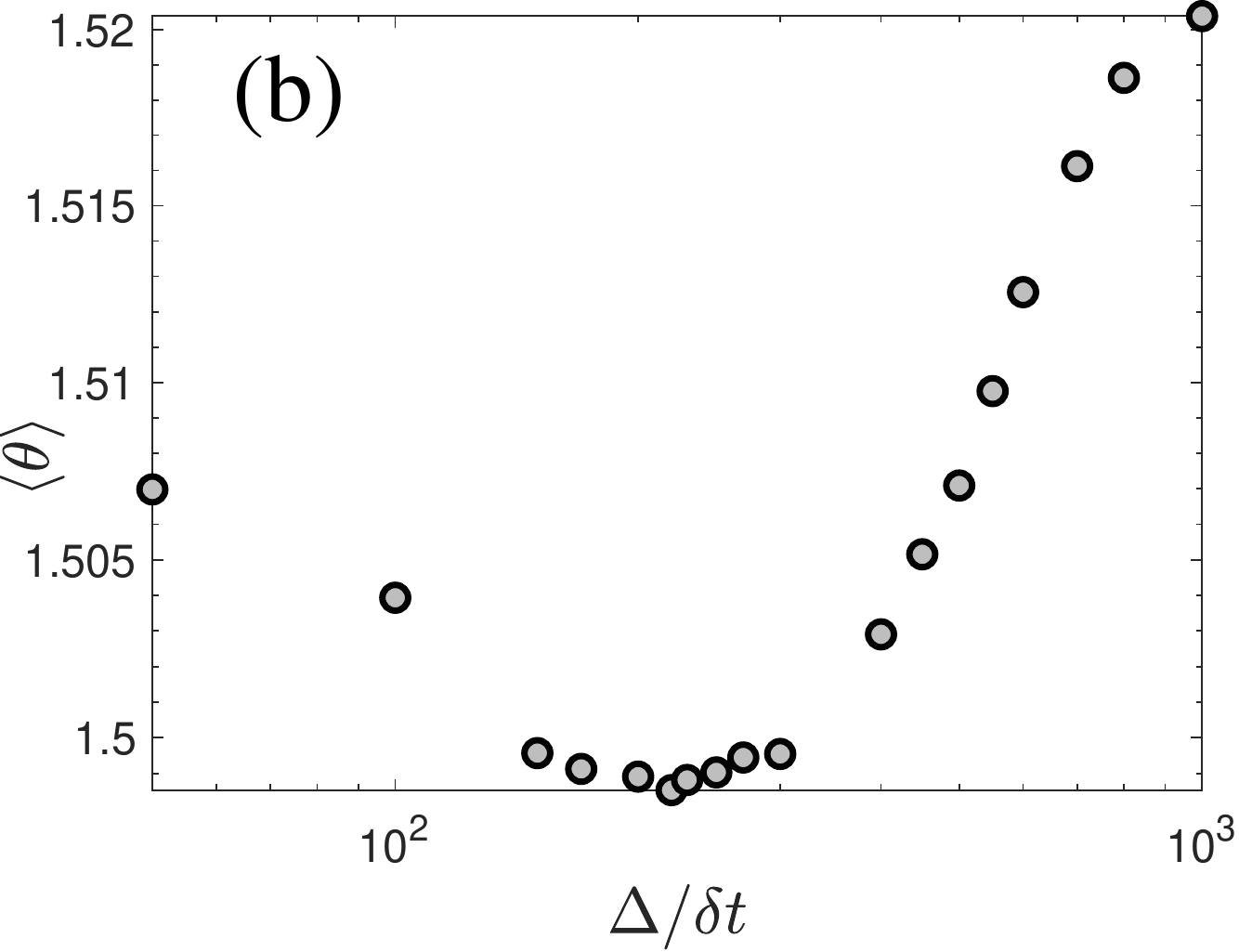} &
\includegraphics[width=60mm]{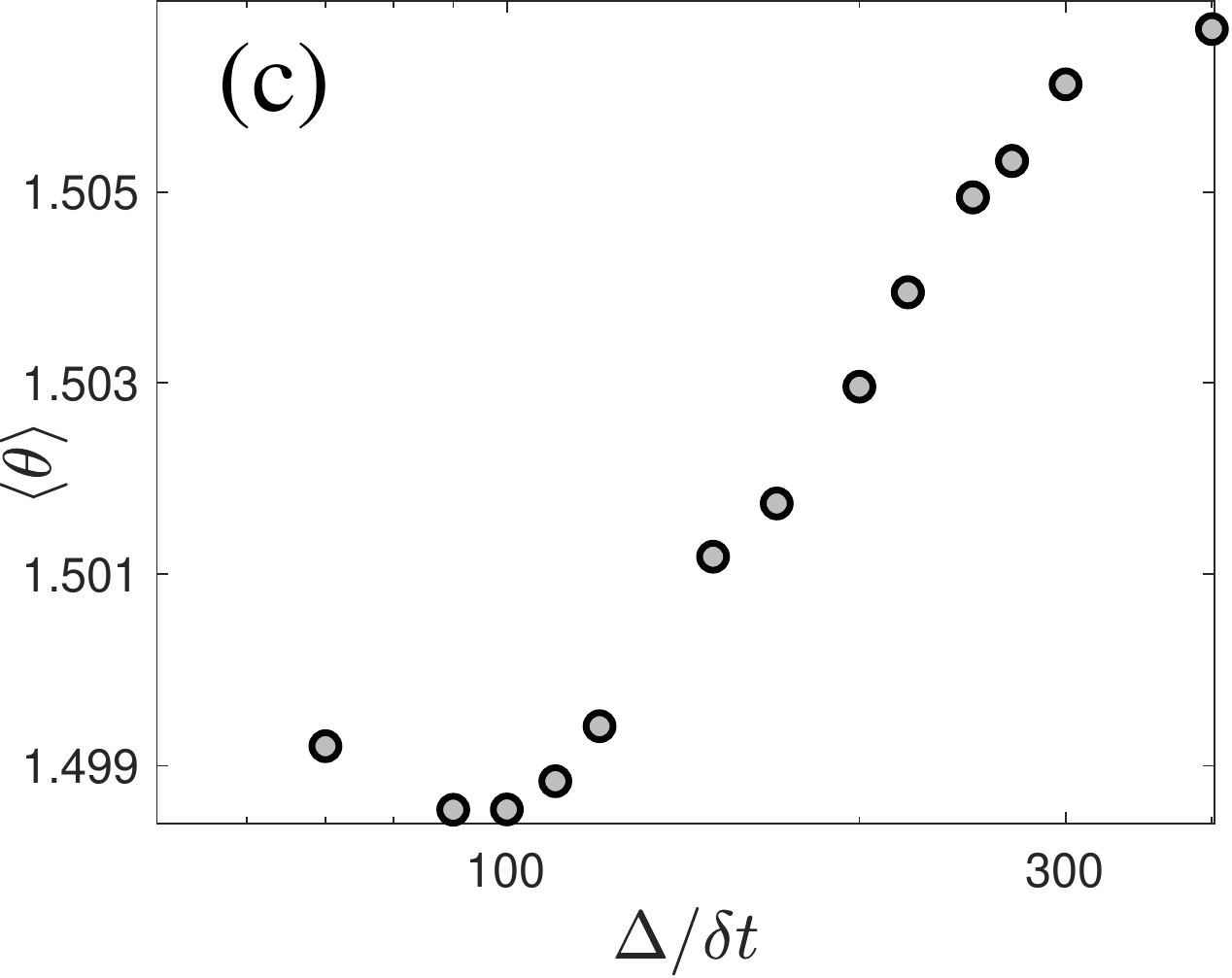}
\end{array}$
\end{center}
\caption{
Relative angle $\langle \theta \rangle$ as a function of time-span $\Delta/\delta t$ for
 \textbf {a} $\phi=0.3801$
\textbf{b} $\phi=0.4276$,
\textbf {c} $\phi=0.4752$
For all panels, $D_a=0.1$, $D_a=0.012$ $6\times 10^5$ simulation steps were performed.  Averaging  over time and $40$ ensembles was performed.
} 
\label{fig:H_suplemental}
\end{figure}

In the main text Figure 2 {\bf(c)} presents the transition from passive to directed motion. 
The order parameter that we use is $\pi/2-\langle \theta \rangle _{min}$. When $\langle\theta\rangle_{min}$ is obtained from searching the minimal value of the average relative angle $\langle \theta \rangle$ as a function of $\Delta$. 
While Fig.~2 {\bf(b)} displays the evolution of $\langle \theta \rangle$ for specific ratios $D_b/D_a$ and the solution density $\phi$, in Fig.~2 {\bf(c)} additional information was incorporated. 
For any given value of the ratio $D_b/D_a$ we plotted the evolution of $\langle \theta \rangle$ as a function $\Delta$ for many values of $\phi$, as presented in Fig.~\ref{fig:H_suplemental} panels {\bf(a)-(c)}. 
For each value $\phi$ a single $\langle \theta\rangle_{min}$ was recorded and then we choose the minimal value of $\langle\theta\rangle_{min}$ that was recorded for all explored values of $\phi$. 
Only this single minimal value of $\langle \theta \rangle_{min}$ is incorporated in Fig.~2 {\bf(c)} as a single point of $\pi/2-\langle \theta \rangle_{min}$ for a given value of $D_b/D_a$.

\section{Possible Mechanisms of Diffusion Coefficient Enhancement}

The enhancement of the diffusion coefficient can occur for various reasons and this is why it is important to elucidate the microscopical scenario that leads to $D_b^\infty$ enhancement. 
In this section, we provide a simple example of why several possible scenarios can potentially enhance the diffusion coefficient.

We use a $1$-dimensional overdamped Langevin equation to describe the position of a single  particle
\begin{equation}
\frac{dx(t)}{dt}=\sqrt{2D(t)}\zeta_t,
    \label{eq:enhance01}
\end{equation}
where $\zeta_t$ is Gaussian $\delta$ correlated random noise and $D(t)$ is a process that defines the diffusion coefficient of the particle. For the simple case when $D(t)$ is constant, i.e., $D(t)=D_0$ $\forall t$, a regular diffusion with diffusion coefficient $D_0$ is obtained
\begin{equation}
\langle x^2(t) \rangle = 2D_0 t.
    \label{eq:enhanceSmpl}
\end{equation}
But for the general case of time-dependent $D(t)$ E.~\eqref{eq:enhanceSmpl} is not valid. Instead, we obtain for $x^2(t)$
\begin{equation}
\langle x^2(t) \rangle = 2\int_0^t D(t')\,dt'
    \label{eq:enhanceCorr}
\end{equation}
and the integral $\int_0^t D(t')\,dt'$ defines the behavior of the second moment. 
The enhancement of the diffusion coefficient in the case of Eq.~\eqref{eq:enhance01} can be obtained simply due to the properties of the process $D(t)$. 
Any process that for short times ($t\to 0$) converges to a constant value $D_0$ and for long times ($t\to\infty$) converges to a larger value $D_1>D_0$ will produce the effect observed in Fig.~1 {\bf(a)} of the main text. 
Such enhancement is also present for the MSD of active particles (see the first Section above).
For example a process $D(t)$ that is a random dichotomous process, i.e. $D(t)$ jumps between two values $D_0$ and $D_1$~\cite{Mario2021}. 
If for short times there is a preference for simple diffusive motion with diffusion coefficient $D_0$ and for longer times the preference is for (again) simple diffusion motion with $D_1(>D_0)$ an enhancement of the diffusion coefficient (as a function of time) will be observed.
For both values of the process $D(t)$ the particle will perform regular Brownian motion without any memory. The enhancement occurs due to the properties of $D(t)$. 
Such behavior of $D(t)$ can possibly mimic  the situation when the particle from time to time experiences strong interactions with other particles in the suspension that lead to strong ``bombardments" of the particle and large uncorrelated excursions, i.e. diffusion with coefficient $D_1$. 
When no interactions are present, the particle simply diffuses with a diffusion coefficient $D_0$. 
No temporal directional preference is present in this model and still, the enhancement of the diffusion coefficient will be observed. 
Therefore proper exploration of the microscopic behavior that leads to enhancement is needed.

\section{Software}
The simulation code can be found in GitLab, project ID 38463829. For technical issues please contact Deborah Schwarcz (deborah.schwarcz@gmail.com).   

\bibliography{./deborah.bib}